\documentclass[useAMS,usenatbib]{mn2e}
\usepackage[british]{babel}	
\usepackage{amsmath}
\usepackage[T1]{fontenc}
\usepackage{verbatim}
\usepackage{graphicx,subfigure,float}
\usepackage{txfonts}
\usepackage[applemac]{inputenc} 
\usepackage{upgreek}
\usepackage{hyperref}

\usepackage{tikz}
\usetikzlibrary{calc,trees,positioning,arrows,chains,shapes.geometric,%
    decorations.pathreplacing,decorations.pathmorphing,shapes,%
    matrix,shapes.symbols}

\usepackage[a4paper]{geometry}
\skip\footins=19.5pt plus 12.0pt minus 1.0pt
\def\hi{\ifmmode{\rm HI}\else{H\/{\sc i}}\fi} 
\newcommand {\kms} {\,{\rm km\,s}^{-1}}
\newcommand {\pc} {\,{\rm pc}}
\newcommand {\kpc} {\,{\rm kpc}}

\newcommand {\de}{^{\circ}}

\newcommand {\ha}{{\rm H}\upalpha}

\newcommand{\bba}{$^{\scriptstyle 3\mathrm{D}}$B{\sc arolo}}
\newcommand{\bbabold}{$^{\scriptstyle \mathbf{3D}}$B{\sc\bfseries arolo}}
\newcommand{\rotcur}{R{\sc otcur}}
\newcommand{\abs}[1]{\lvert#1\rvert}
\newcommand{\vir}[1]{``#1''}
\newcommand {\vrot}{{V_\mathrm{rot}}}
\newcommand {\vsys}{{V_\mathrm{sys}}}
\newcommand {\vdisp}{{\sigma_\mathrm{gas}}}

\newcommand{\reffig}[1]{Fig. \ref{#1}}
\newcommand{\refsec}[1]{Section \ref{#1}}

\title[{\emph{\bba}}: a new 3D algorithm to derive rotation curves of galaxies]{\bbabold: a new 3D algorithm to derive rotation curves of galaxies\thanks{\bba\ is available for download at \url{http://editeodoro.github.io/Bbarolo} }}
\author[E. M. Di Teodoro and F. Fraternali]{E. M. Di Teodoro$^{1}$ and F. Fraternali$^{1,2}$\\
$^{1}$Department of Physics and Astronomy, University of Bologna, 6/2, Viale Berti Pichat, 40127 Bologna, Italy\\
$^{2}$Kapteyn Astronomical Institute, Postbus 800, 9700 AV Groningen, The Netherlands}

\begin{document}


\pagerange{\pageref{firstpage}--\pageref{lastpage}} \pubyear{2015}

\maketitle

\label{firstpage}

\begin{abstract}

We present \bba, a new code that derives rotation curves of galaxies from emission-line observations. This software fits 3D tilted-ring models to spectroscopic data-cubes and can be used with a variety of observations: from \hi\ and molecular lines to optical/IR recombination lines. We describe the structure of the main algorithm and show that it performs much better than the standard 2D approach on velocity fields. A number of successful applications, from high to very low spatial resolution data are presented and discussed. \bba\ can recover the true rotation curve and estimate the intrinsic velocity dispersion even in barely resolved galaxies ($\sim$2 resolution elements) provided that the signal to noise of the data is larger that 2-3. It can also be run automatically thanks to its source-detection and first-estimate modules, which make it suitable for the analysis of large 3D datasets. These features make \bba\ a uniquely useful tool to derive reliable kinematics for both local and high-redshift galaxies from a variety of different instruments including the new-generation IFUs, ALMA and the SKA pathfinders.\\

\end{abstract}


\section{Introduction}

The study of the gas kinematics is one of the most powerful tools to infer the distribution of dark and luminous matter and understand the dynamical structure of disc galaxies. In this context, spectroscopy plays a primary role since it can naturally trace the motion of matter thanks to the Doppler shift of the emission lines. 
In particular, observations of the neutral hydrogen (\hi) emission-line \citep*[e.g.][]{Rogstad&Shostak71, Allen+73} allow to analyze the kinematics of galaxies and derive rotation curves out to very large radii. 
The total mass and the gravitational potential can be inferred by decomposing the rotation curve into the contribution given by the different matter components in a galaxy, i.e.\ the gas, the stars and the dark matter (DM)  \citep[e.g.][]{vanAlbada+85}. Observed rotation curves broadly have two different shapes \citep[e.g.][]{Casertano&vanGorkom91}: large spiral galaxies show steeply rising rotation curves that quickly become and stay flat in the external regions. Dwarf galaxies have slowly rising rotation curves that may not reach a flat part. The former are believed to be dominated by baryons in the inner regions and by DM in the outer regions. The latter are thought to be DM-dominated at all radii and they have been largely used to constrain the inner slope of the density profile of DM halos \citep[e.g.][]{KuziodeNaray+08, vanEymeren+09}. 

Rotation curves are also fundamental to study the most important scaling relations for spiral galaxies: the \vir{Tully-Fisher Relation} \citep[TFR,][]{Tully&Fisher77} and the \vir{Baryonic TFR} \citep{McGaugh+00}, correlating the rotation velocity (hence the dynamical mass) to the stellar luminosity and to the total baryonic mass, respectively. The TFR is a severe testing ground for galaxy formation: current observations show a slope $L$ (or $M_\mathrm{bar}$) $\propto V^{\alpha}$  with $\alpha\sim4$ \citep[e.g.][]{Verheijen01,McGaugh12} and these relations strongly constrain models in the standard $\Lambda$CDM picture \citep[e.g.][]{Mo+98,Bullock+01}. Finally, gas kinematics can also be used to infer the distribution of angular momentum in galaxies and study how it evolves with time \citep[e.g.][]{Romanowsky&Fall12}.

All these fields of study require a precise determination of the rotation curves, from the innermost to the outermost parts of galaxy discs. The most common approach to describe the kinematics of spiral galaxies make use of the so-called \vir{tilted-ring model} \citep*[e.g.][]{Rogstad+74}. Such a model is based on the assumption that the emitting material is confined to a thin disc and that the kinematics is dominated by the rotational motion, so that each ring has a constant circular velocity $\vrot(R)$, depending only on the distance $R$ from the centre. This approximation is very good for local disc galaxies without bars as their orbits are nearly circular, with a typical axis ratio $b/a>0.9$ \citep*[e.g.][]{Franx&deZeeuw92, Schoenmakers+97}. In a tilted-ring model, the disc is therefore broken down into a number of concentric rings with different radii, inclinations, position angles and rotation velocities. 

The standard approach to link the tilted-ring model to observations is through the fitting of 2D velocity fields. The observed velocities along the rings can be written as a finite number of harmonic terms \citep*[e.g.][]{Franx+94}. At the first order and neglecting the streaming motions, this expansion leads to the expression $V_\mathrm{los}(R) = \vsys +\vrot(R)\cos\theta\sin i$, where $\vsys$ is the systemic velocity and $\theta$ is the azimuthal angle, measured in the plane of the galaxy ($\theta=0$ for major axis), related to the inclination $i$ and the position angle $\phi$, measured on the plane of the sky (see \reffig{fig:disc}). It is then straightforward to fit such a function to a velocity field by using any of the numerous non-linear least-squares fitting techniques \citep[e.g.][Chap. 15]{Press+07}. There are several available algorithms that perform this fit to velocity fields. The most used is \rotcur\ \citep{vanAlbada+85, Begeman87}, which can be found in data analysis packages like AIPS \citep{Fomalont81}, GIPSY \citep{vanderHulst+92} and NEMO \citep{Teuben95}. \rotcur\ simply fits the  above function by using a Levenberg-Marquardt solver \citep{Levenberg44, Marquardt63}. An optional term for the expansion/contraction motions through the disc is also available in \rotcur. In the last ten years, several new codes have refined this basic 2D approach by experimenting different fitting techniques and by extending the harmonic expansion to higher-order terms to take into account asymmetries and non-circular motions. These 2D fitting algorithms include \textsc{Reswri} \citep{Schoenmakers99}, \textsc{Ringfit} \citep{Simon+03}, \textsc{Kinemetry} \citep{Krajnovic+06} and \textsc{DiskFit} \citep{Spekkens&Sellwood07, Sellwood&Sanchez10}.

All the above 2D algorithms are fast from a computational point of view and do return good kinematic models and reliable rotation curves when applied to high-resolution velocity fields. A number of drawbacks however exist. First of all, a velocity field is not the full data set, but it is itself derived from a data-cube, requiring the intermediate step of extracting a characteristic velocity from the line profile at each spatial pixel. Most used approaches include intensity-weighted mean along the velocity axis \citep[1$^\mathrm{st}$ moment,][]{Rogstad&Shostak71}, fitting single \citep{Begeman87, Swaters99} or multiple \citep{Oh+08} Gaussian functions to the profiles, and/or including a skewness term \citep[e.g.\ $h_3$ Gauss-Hermite polynomial,][]{vanderMarel&Franx93}. Unfortunately velocity fields derived with different methods can significantly deviate from each other, especially for galaxies with asymmetric profiles. Moreover, the derivation of an unambiguous velocity is not possible when the line of sight with respect to the observer intersects the disc twice or more, like in nearly edge-on systems or in the presence of thick discs and outer flares.

The most severe problem in deriving the kinematics from the velocity field is however the beam smearing  \citep[e.g.][]{Bosma78,Begeman87}. The finite beam size of a telescope causes the line emission to be smeared on the adjacent regions. The effect is that the gradients in the velocity fields tend to become flatter while the line-profiles in each spatial pixel become broader. In other words, part of the rotation velocity is turned into line broadening that can be erroneously interpreted as gas velocity dispersion, producing a well-known degeneracy between these two quantities (see \refsec{sec:verylow}). The typical effect of the beam smearing is that the derived rotation curves will rise very slowly in the inner regions with potentially dramatic consequences for the interpretation \citep[e.g.][]{Lelli+10}. The effect becomes more and more pronounced as the spatial resolution of the observations decreases and the inclination angle of the galaxy increases.

The natural solution for almost all the above-mentioned issues is a 3D modeling of the data-cubes. A 3D approach gives larger opportunities for modeling galaxies and is effectively not affected by the beam smearing, since the instrumental effect is introduced in the model through a convolution step (see \refsec{sec:conv}). The main drawback is the computational slowness. Unlike the 2D tilted-ring model, an analytic form for the fitting function in 3D does not exist and the model is instead constructed by a Monte-Carlo extraction (see \refsec{sec:discmodel}). Thus, the fit must be performed with algorithms that do not require the knowledge of any partial derivative \citep[e.g.][Chap. 10]{Press+07}. Such techniques are known to be computationally expensive and they may converge to a local minimum of the function. In addition, the minimization is not just performed on a single map, but on the whole data-cube, which consists of $n$ maps, where $n$ is the number of channels. Finally, a larger number of parameters than in 2D is needed to describe the model.

The visual comparison between the data-cube and an artificial model cube has been used as an additional step to improve the results of the 2D tilted-ring model and potentially correct for the beam smearing, especially in the investigation of the shape of dark matter profiles in the inner region of galaxies \citep[e.g.][]{Swaters99,Gentile+04}. A pioneering attempt to directly use a 3D approach was made by \cite{Corbelli&Schneider97} in a study on the outer warp of M33. A  currently available algorithm that can directly fit a 3D tilted-ring model to data-cubes is TiRiFiC \citep{Jozsa+07}. TiRiFiC has reached a considerable degree of development and sophistication and it has been successfully used to study the kinematics of nearby galaxies with peculiar features, like strong warps and extra-planar gas \citep{Jozsa+09, Zschaechner+12, Gentile+13}. A 3D approach is also used in a recent tool \citep[GalPak3D,][]{Bouche+15} targeted to high-redshift galaxies.

In this paper, we present a new software to automatically fit 3D tilted-ring models to emission-line data-cubes. The name of the code is \bba, which stands for \vir{3D-Based Analysis of Rotating Objects from Line Observations}. This code works with 3D FITS images having two spatial dimensions and one spectral dimension (velocity, frequency or wavelength). \bba\ builds a number of models in the form of artificial 3D observations and compares them with the input cube, finding the set of geometrical and kinematical parameters that better describes the data. Our purpose is to provide an easy-to-use algorithm that might be applicable to a wide range of emission-line observations, from radio-\hi\ data of local galaxies to sub-mm and optical/IR lines of galaxies up to high redshift, from high to very low spatial resolution. Unlike TiRiFiC, which has been mainly developed to study local galaxies with a detailed description of peculiarities, such as warps, spiral arms and lopsidedness, \bba\ is designed to work on low-resolution data, where the kinematic information is largely biased by the size of the beam. Therefore, we kept the disc model as simple as possible and we restrain the number of parameters to the bare minimum. 

This paper is organized as follows. In \refsec{sec:algorithm} we illustrate \bba 's main algorithm, going through the main fitting steps and describing their features. In \refsec{sec:tests} we show some applications and tests with \hi\ data. We start with the modeling of a well-known local galaxy at high spatial resolution and then move to lower resolution data. We discuss the robustness and the limits of \bba\ and we show the differences between a traditional 2D and our 3D approach in data largely afflicted by beam smearing. In \refsec{sec:mock} we test the accuracy of the code by using mock galaxies. \refsec{sec:conclusions} recaps and discusses possible developments and future applications of \bba.

\section{The main algorithm}
\label{sec:algorithm}

\begin{figure}
\centering
\tikzstyle{block} = [rectangle, rounded corners, fill=orange!30,draw=black, very thick,text width=7em, minimum height=3em, text centered]
   
\tikzstyle{decision} = [diamond, draw=black, very thick, fill=olive!10,
    text width=6em, text badly centered, inner sep=0pt, aspect=1.6]
    
\tikzstyle{io} = [trapezium, trapezium left angle=70, trapezium right angle=110, draw=black, very thick ,fill=teal!15, minimum height=3em, minimum width=7em,text badly centered]

\tikzstyle{cloud} = [draw, ellipse,fill=cyan!20, node distance=2.5cm, draw=black, very thick,minimum height=3em, minimum width=4em, text badly centered]
    
\tikzstyle{line} = [draw, very thick, color=black!55, -latex']

\begin{tikzpicture}[scale=2, node distance = 1.5cm, auto,->=stealth]

    \node [io] (initial) {Input};
    \node [decision, below of=initial,node distance=1.6cm] (decidering) {R<R$_\mathrm{max}$?};
    \node [block, below of = decidering, node distance=1.6cm] (model) {Disc model};
    \node [block, below of=model] (smooth) {Convolution};
    \node [block, below of=smooth] (resid) {Residuals};
    \node [decision, below of=resid, node distance=1.6cm] (decide) {Converges?};
    \node [cloud, below left=1.2cm and 0.8cm of  model,text width=4.5em] (update) {Update parameters};  
    \node [cloud, below right=2.3cm and 1.4cm of decidering,text width=4em] (updateR) {R=R+$\Delta$R};  
    \node [io,left=1.cm of decidering, text width=3em] (final) {Final model };

     \path [line] (initial) -- node [, color=black] {R=R$_\mathrm{min}$} (decidering);
     \path [line] (decidering) -- node [, color=black] {yes} (model);
     \path [line] (decidering) -- node [, color=black] {no} (final);
     \path [line] (updateR) |- (decidering);
     \path [line] (model) -- (smooth);
     \path [line] (smooth) -- (resid);
     \path [line] (resid) -- (decide); 
  	 \path [line] (decide) -| node [near start, above, color=black] {no} (update);
  	 \path [line] (decide) -| node [near start, above, color=black] {yes} (updateR);
     \path [line] (update) |- (model);

\end{tikzpicture}

\vspace{0.465cm}
\caption{Flowchart of the \bba\ main algorithm. For each ring R, the algorithm builds a 3D model, convolves it with the observational beam/PSF and compares it with the data. If no convergence has been achieved, \bba\ updates the parameters and starts over. When the algorithm converges to the minimum, it moves to the next ring. The optional normalization step takes place after the convolution step. }
\label{fig:flowchart}
\end{figure}
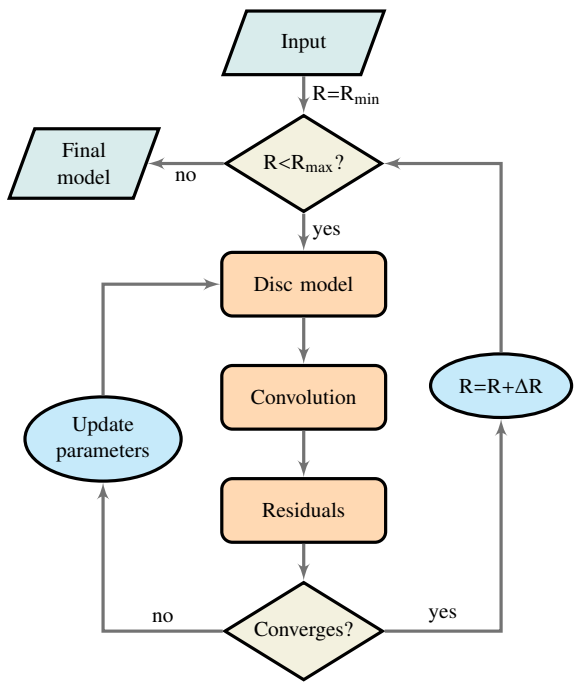

The main feature of \bba\ is that it simulates data-cube observations, starting from the model of a rotating gaseous disc, and compares them with real data. The disc is made up by a number of concentric rings with non-zero thickness. The emission from gas in each ring is generated in a 6D domain (three dimensions for the spatial location and three for the components of the velocity) and these rings are then projected into the 3D data-cube (i.e.\, two spatial and one spectral dimensions). The comparison with the data is perfomed ring by ring. At each step, if the model is good enough (see below), the algorithm moves to the following ring, otherwise it updates the disc parameters until the best match between model and observations is found.

The minimization between model and data is performed by using the multidimensional downhill simplex solver, also known as the Nelder-Mead method \citep{Nelder&Mead65} for the minimization of non-analytic functions. 
The user supplies a number of initial guesses for each model ring from which the function to be passed to the minimization algorithm is built. Input parameters can be either globally or ring-by-ring defined. If no guess is given, the algorithm will autonomously estimate the initial parameters for the fit (see \refsec{sec:addfeat}). 
The function to minimize is an indicator of how the model and the data differ from each other and its construction occurs through four main steps:

\begin{enumerate}
\item \emph{\large Disc model.} The disc model is built by a Monte-Carlo reproduction of the gas distribution both in the space and velocity domain. This function derives from the \textsc{Galmod} routine \citep{Sicking97} implemented in GIPSY. \vspace{4pt}

\item \emph{\large Convolution.} The model is degraded to the same spatial resolution of the data via the convolution with a 2D Gaussian representing the observational Point Spread Function (PSF). The spectral broadening is instead taken into account in the model construction (see \refsec{sec:discmodel}) \vspace{4pt}

\item \emph{\large Normalization.} The model is normalized to the 0th moment map of the observations pixel-by-pixel or azimuthally. This step can be skipped and the density profile supplied by the user or possibly fitted. \vspace{4pt}

\item \emph{\large Residuals.} Comparison between the model and the data pixel-by-pixel. The sum of the residuals is returned to the minimizing algorithm.
\end{enumerate}

A schematic flowchart of the main fitting algorithm is shown in \reffig{fig:flowchart}. In the following sections, we describe the most important steps and the main features of the algorithm.

\subsection{Disc model}
\label{sec:discmodel}

The artificial gaseous disc is constructed with a 3D tilted-ring model by using a stochastic function that randomly populates the space with emitting gas clouds, for which line profiles are built. Each ring of radius $R$ and width $W$, is described by the following geometrical and kinematical parameters:

\begin{itemize}
\item[-] Spatial coordinates of the centre ($x_0$, $y_0$).
\item[-] Systemic velocity $\vsys$.
\item[-] Inclination angle $i$ w.r.t. the observer (90$\de$ for edge-on).
\item[-] Position angle $\phi$ of the major axis on the receding half of the galaxy, taken anticlockwise from the North direction on the sky.
\item[-] Rotational velocity $\vrot$.
\item[-] Velocity dispersion $\vdisp$.
\item[-] Face-on gas column density $\Sigma$.
\item[-] Scale-height of the gas layer $z_0$.
\end{itemize}

\begin{figure}
\centering
\includegraphics[width=0.5\textwidth]{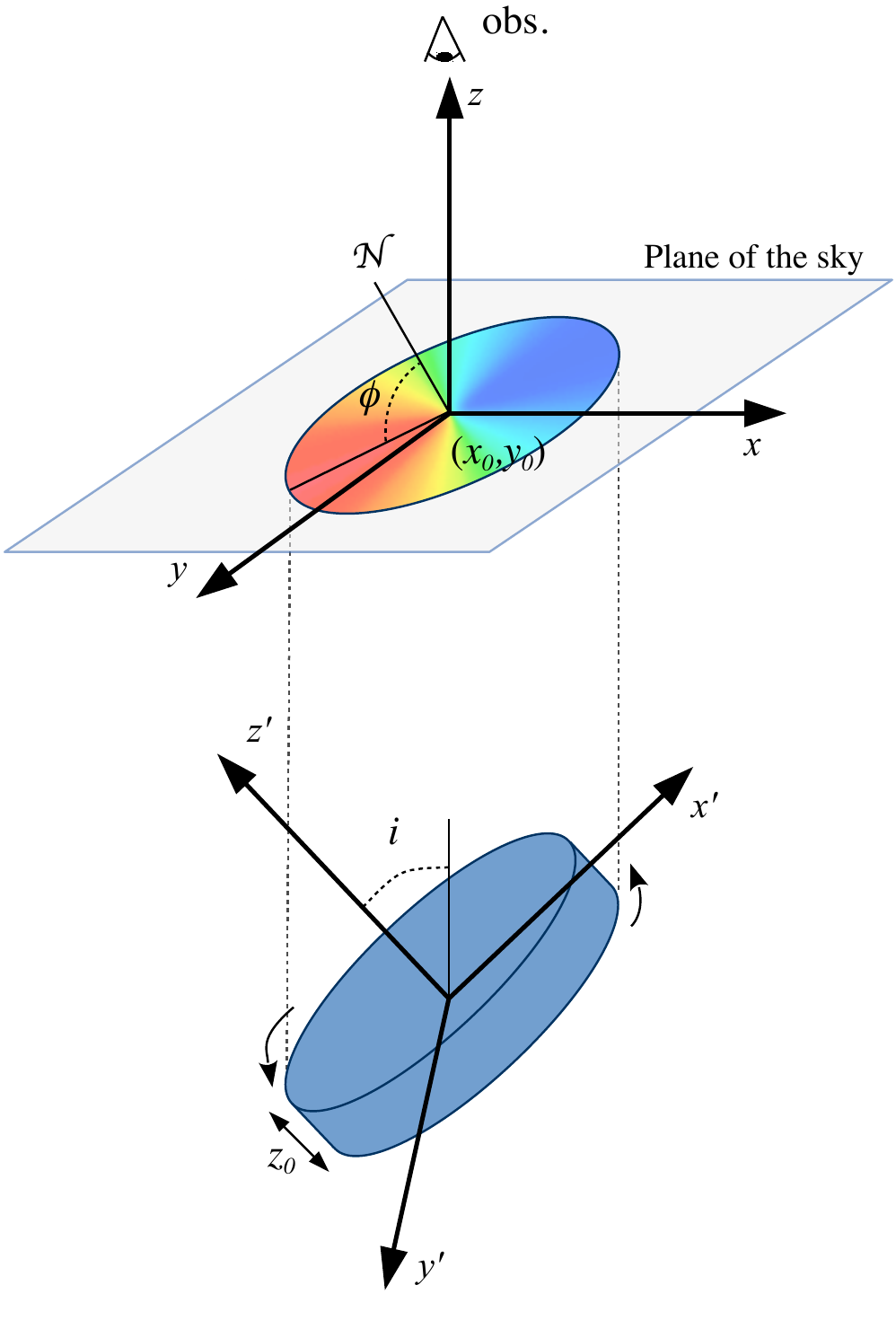}
\caption{Geometrical parameters of the disc model. The galaxy disc in the $x'y'z'$ space is projected into an ellipse in the $xy$ plane of the sky. The inclination angle $i$ is taken with respect to the line of sight, the position angle $\phi$ identifies the position of the major axis on the receding half of the galaxy and it is taken counterclockwise from the North direction.}
\label{fig:disc}
\end{figure}

All these quantities are allowed to vary from ring to ring. The first six parameters are the same required by 2D fitting algorithms like \rotcur .  The geometry of the tilted-ring model is shown in \reffig{fig:disc}.
Each ring is filled with gas clouds whose spatial position is given in cylindrical coordinates by a radius $R_\mathrm{c}$ (with \textit{R-W/2 < $R_\mathrm{c}$ < R+W/2}), an azimuthal angle $\theta_\mathrm{c}$  ($0\leq\theta_\mathrm{c}\leq2\pi$) and a height $z_\mathrm{c}$ above the plane of the disc. Radius and azimuth are randomly and uniformly chosen, the height is selected as a random deviate from a given vertical distribution of the gas density (available functions are Gaussian, sech$^2$, exponential, Lorentzian and box layer). The position of the clouds is then rotated and projected onto the plane of the sky with a given orientation with respect to the observer, according to the position angle and inclination at that radius.

Once the positions of the clouds are determined, the observed velocities along the line of sight are calculated as a combination of systemic, rotational and random motions. The velocity profile at each location is built around the average velocity by dividing the clouds into a number of sub-clouds with velocities distributed as a Gaussian with dispersion $\sigma^2 = {\sigma^2_\mathrm{gas}+\sigma^2_\mathrm{instr}}$, being $\vdisp$ the intrinsic gas dispersion and $\sigma_\mathrm{instr}$ the instrumental broadening. These velocities are then discretized and the contribution of the sub-clouds is recorded in a model cube with the same sizes of the data-cube. 
\bba\ uses by default $\sigma_\mathrm{instr}=W_\mathrm{ch}/ \sqrt{2\ln  2}$, where $ W_\mathrm{ch}$ is the channel width of the data-cube.  This is usually a good assumption in \hi\ data-cubes where Hanning smoothing has been applied \citep{Verheijen97}. Otherwise, the user can supply an own value for the spectral resolution.

\subsection{Convolution}
\label{sec:conv}

The model has a nominal spatial resolution of a pixel and it needs to be smoothed to the same spatial resolution of the data. This requires to perform a spatial convolution by the observational PSF, or beam, for each spectral channel. We approximate the PSF as a two-dimensional Gaussian function, which is an adequate choice for radio, millimeter/submillimeter and optical/IR observations.
The Gaussian function is defined by three parameters that characterize the elliptical cross-sectional shape of the kernel: the full-width half-maximum (FWHM) of both the major and minor axes ($a$ and $b$, respectively), and the position angle of the major axis ($\psi$), measured anticlockwise from the vertical direction. The kernel of the two-dimensional Gaussian is defined by the function:

\begin{equation}
	k(x,y) = \frac{1}{2\pi\lambda_\kappa\lambda_\eta} \exp{\left[-0.5 \left( \frac{\kappa^2 (x,y)}{\lambda_\kappa^2} + \frac{\eta^2(x,y)}{\lambda_\eta^2} \right) \right]}
	\label{eq:gausskern}
\end{equation}

\noindent where $(x,y)$ are the offsets from the centre of the Gaussian and $\lambda_\kappa$ and $\lambda_\eta$ are the standard deviations along $(\kappa,\eta)=(x\sin\psi-y\cos\psi, \, x\cos\psi-y\sin\psi)$, the position-angle-rotated frame of reference. The FWHMs ($a=\sqrt{8\ln 2}\,\lambda_\kappa$ and $b=\sqrt{8\ln 2}\,\lambda_\eta$) and the position angle are usually read from the header as the \textsc{bmaj}, \textsc{bmin} and \textsc{bpa} keywords, respectively, but they can also be manually supplied. In addition, for Integral Field Unit (IFU) data, \bba\ can receive as input an image (or data-cube) of a star, usually observed at the same time of the scientific target. The star is then used to determine the PSF by fitting it with a 2D Gaussian.

This smoothing step is the bottleneck of the fitting algorithm, since the convolution is a very computational expensive operation and \bba\ needs to perform it for each calculated model. To speed-up this step with Fast-Fourier transforms, we used the shared-memory parallel OpenMP implementation of the FFTW3 library. The user is however advised to use data-cubes with suitable sampling in order to save computational time.

\subsection{Normalization}
The normalization allows the code to exclude one parameter from the fit, namely the surface density $\Sigma$ of the gas. We have currently implemented two different kinds of normalization: pixel-by-pixel and azimuthally averaged. In the former case, the model is normalized such that the column density maps of model and observations are the same. In other words, we impose that the integral of each spatial pixel along the spectral dimension in the model is equal to the integral of the correspondent spatial pixel in the observations. This type of normalization allows to have a non-axisymmetric model in density and avoids that untypical regions, like areas with strong and clumpy emission or holes, might affect the global fit \citep[see e.g.][]{Lelli+12}. In the second case, the model is instead normalized to the azimuthal-averaged flux in each ring. According to our tests, the pixel-by-pixel normalization is often a more advisable solution, so this is the default in \bba. The azimuthal-averaged normalization is useful to determine the inclination angle of the outer rings. The normalization step can be turned off, in this case the user can supply a surface density profile or leave it free and fit it together with the other parameters.

\subsection{Residuals}

The residuals are calculated by comparing the model and the data pixel-by-pixel.
The number $F$, which is returned to the minimization algorithm and defines whether a model is suitable or not, is the averaged sum of the residuals over each pixel:

\begin{equation}
\label{eq:res}
F=\frac{1}{n} \sum_{i=1}^{n} \Delta r_i \, w(\theta_i)
\end{equation}

\noindent where $n$ is number of pixels where the residual $\Delta r$ are evaluated and $w(\theta)$ is a weighting function. To be considered, a pixel must either have a non-zero flux in the model or be part of the identified emission region of the galaxy. We provide three kinds of residuals:

\begin{subequations}
\begin{equation}\label{eq:chisq}
	\Delta r = \frac{(M-D)^2}{\sqrt D}
\end{equation}
\begin{equation}\label{eq:absdiff}	
\Delta r = \abs{\,M-D\,}
\end{equation}
\begin{equation}\label{eq:reldiff}
\Delta r = \frac{\abs{\,M-D\,}}{(M+D)}
\end{equation}
\end{subequations}

\noindent where $M$ and $D$ are the flux values of the model and the data, respectively. The (\ref{eq:chisq}) residual  is a kind of $\chi^2$ without however a conventional statistical meaning. When $D$ is a blank pixel, in (\ref{eq:chisq}) we set $D$ equal to the root mean square value (rms) of the cube. The (\ref{eq:reldiff}) residuals give more weight to regions  where the emission is faint and diffuse, the (\ref{eq:absdiff}) is intermediate. The weighting function in Eq. (\ref{eq:res}) is $w(\theta)=\abs{\cos \theta}^m$, where $\theta$ is the azimuthal angle (0$\de$ for the major axis) and $m=0,1,2$. With $m\neq0$, the residual gives prominence to regions close to the major axis, i.e.\ where most of the information on the rotation motions lies. As an option, Eq. (\ref{eq:res}) can be multiplied by a factor $(1+n_\mathrm{b})^p$, where $n_\mathrm{b}$ is the number of pixels having the model but not the observation, in order to penalise models that extend farther than the data. This option is useful, for instance, to determine the correct inclination of the outer rings. As mentioned, this choice should be combined with an azimuthal normalization of the surface density.

\begin{figure*}
\centering
\includegraphics[width=\textwidth]{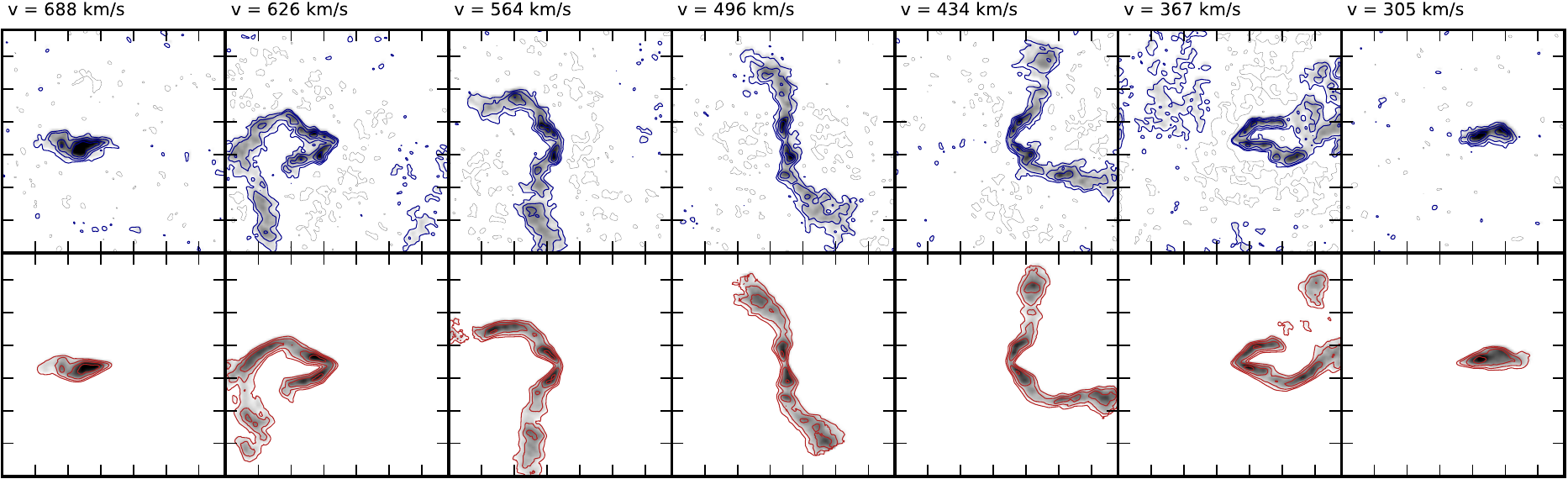}
\caption{Comparison between the observations (blue, \emph{top panels}) and the model (red, \emph{bottom panels}) for NGC 5055 from the THINGS survey. Here we show some representative channel maps taken from a 30" smoothed data-cube. The size of the field is 25$'$ x 25$'$. The lower contours are at 2.5$\sigma$ level, negative contours are shown in grey. The map corresponding to the systemic velocity of the galaxy is shown in the centre (496 $\kms$). Despite the observations have an uneven noise distribution, \bba\ is able to reproduce the data very well and in particular it traces the prominent warp visible as a deformation of the channel maps in the outer regions of the galaxy.}
\label{fig:chanmap}
\end{figure*}

\subsection{Additional features}
\label{sec:addfeat}

Here, we list and briefly describe some other useful tools available inside \bba:

\begin{itemize}
\item[-] \emph{Source detection.} A source finder derived from Duchamp \citep{Whiting12} is implemented. \bba\ can identify all the sources in the given data-cube and automatically fit each of them.\vspace{4pt}

\item[-] \emph{Masking.} In order to obtain a good fit, \bba\ needs to build a mask to identify the regions that are ascribable to the galaxies in the data-cube. The default algorithm takes advantage of the source finder and builds a mask directly on the identified emission regions. As an alternative, \bba\ can smooth the original data-cube by a factor $n$ and consider only the regions where the flux is higher than $m$ times the rms of the smoothed data-cube. Default values are $n=2$ and $m=3$, but they can be changed by the user.\vspace{4pt}

\item[-] \emph{Automatic initial guesses.} \bba\ can automatically estimate the initial parameters for the fit. The algorithm starts by isolating the galaxy through the source finder. The coordinates of the centre are taken as the flux-weighted average positions of the source. The systemic velocity is calculated as the midpoint between the velocities corresponding to the 20\% of the peaks of the global line-profile. The position angle is estimated from the velocity field as the straight line that maximizes the gradient in velocity along the line of sight. The inclination is calculated from the column density map: a model map is built from the observed gas density profile, smoothed to the same spatial resolution of the data and fitted to the observed map. The rotation velocity is calculated as the inclination-corrected half-width of the global line profile at 20\% of the peak flux ($W_{20}$). Default values for the velocity dispersion and the disc thickness are 8 $\kms$ and 150 pc, respectively. The algorithm is able to determine good initial guesses in most cases. The weakest step is the estimate of the inclination, which is also the most critical parameter to fit \citep[see also][]{Begeman87}. These initial estimates allow \bba\ to be run automatically, but the user can also decide to set some parameters and let the code estimate the others. A number of output plots (e.g.\ position-velocity diagrams along the major and minor axes) are provided to allow the user to check the quality of the automatic fits. \vspace{4pt}

\item[-] \emph{Regularization of the parameters.} Fitting the geometrical parameters together with the kinematic ones can lead to unphysical discontinuities in the derived rotation curve. The fits of the inclination and the position angle, in particular, often show unrealistic oscillations and numerical scatter. The usual approach for dealing with this issue, also in the 2D tilted-ring model, consists in deriving a first model by fitting all the parameters together, and a second model by fixing the geometrical parameters to some functional form and fitting only the circular velocity. \bba\ can automatically perform this parameter regularization and then proceed to a second fitting step, with only rotation velocity and velocity dispersion left free. The other parameters are interpolated and fixed either to a polynomial of degree $m$ (chosen by the user, default $m=3$) or a Bezier function. \vspace{4pt}

\item[-] \emph{Errors.} There is no direct way to calculate the errors on the fitted parameters in the 3D approach. We estimate errors via a Monte-Carlo method. Once the minimization algorithm converges, \bba\ calculates a number of models by changing the parameters with random Gaussian draws centred on the minimum of the function. This allows the code to oversample the parameters space close to the minimum. In this region, the residuals usually have the behavior of a quadratic function. Errors for each parameter are taken as the range where this quadratic function shows a residuals increase of a $n$ percentage with respect to the minimum (default value is 5\%). Albeit this procedure is not optimal and computationally expensive, it returns errors which are in good agreement with those obtained with more standard methods, for example the difference between the receding and the approaching halves of the disc \citep[e.g.][]{Swaters99}. 

\end{itemize}

\section{Testing \bbabold}
\label{sec:tests}

\bba\ has been already tested on about one hundred galaxies, especially local systems observed in the \hi-line, but also IFUs (SINFONI and KMOS) data-cubes of high-z systems (Di Teodoro et al., in prep). In this section, we show some applications of \bba\ to \hi\ data-cubes at different spatial resolutions and we compare our results with the traditional 2D approach. 

\subsection{High-resolution data}
\label{sec:hires}

We run \bba\ on several high-resolution galaxies from the available \hi-survey, like The HI Nearby Galaxy Survey \citep[THINGS,][] {Walter+08}, the Very Large Array - ACS Nearby Galaxy Survey Treasury \citep[VLA-ANGST,][]{Ott+12} and the Hydrogen Accretion in LOcal GAlaxies Survey \citep[HALOGAS,][]{Heald+11}. \bba\ has been able to model these high-resolution galaxies and the derived rotation curves agree very well with the already published 2D rotation curves. 

\begin{figure*}
\centering
\includegraphics[width=\textwidth]{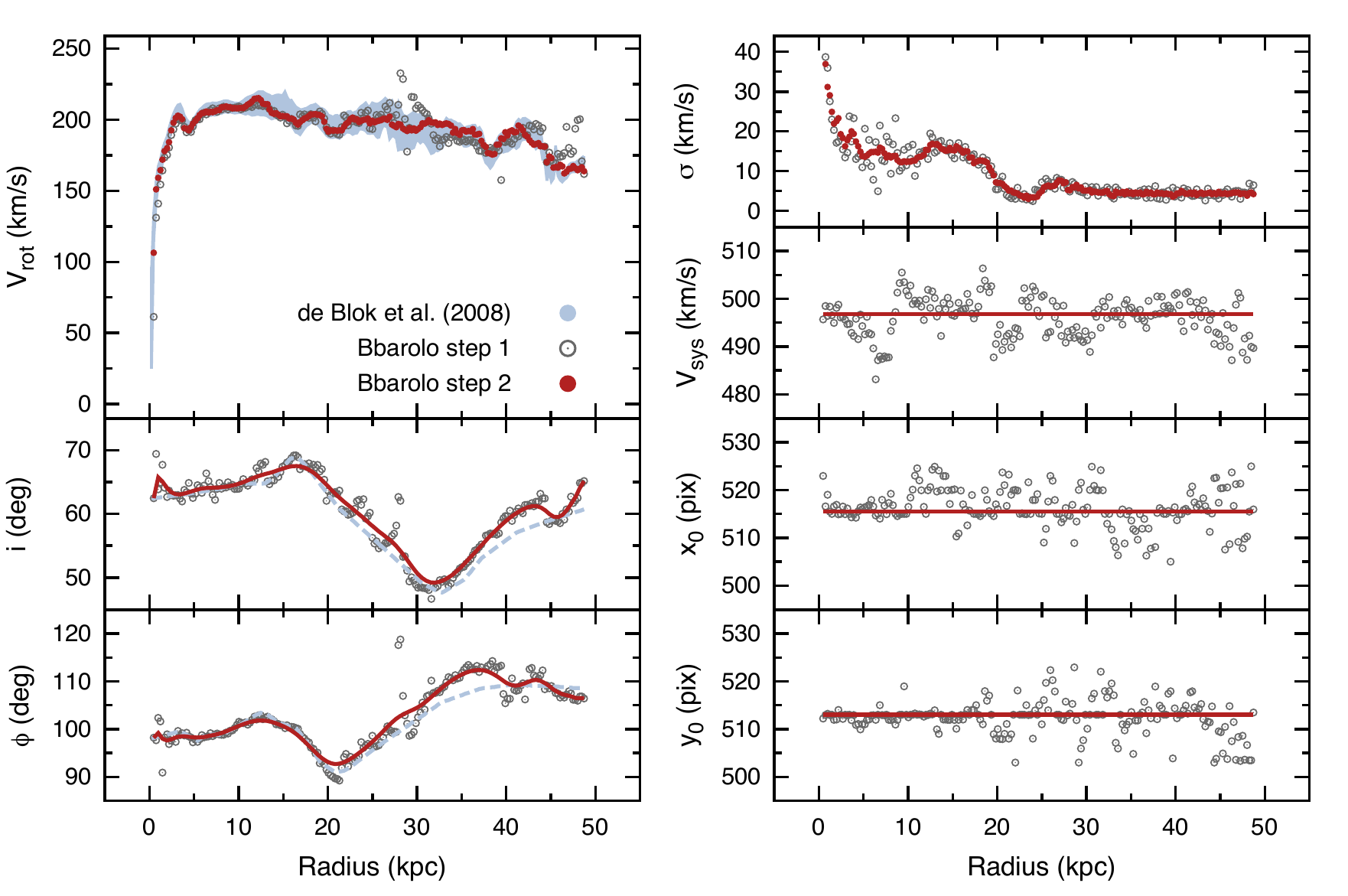}
\caption{Tilted-ring parameters derived with \bba\ using the THINGS data-cube of the warped galaxy NGC 5055. \emph{Left panels}: from the top, the rotation curve, the inclination and the position angle for each ring. \emph{Right panels}: from the top, the velocity dispersion, the systemic velocity and the coordinates of the centre for each ring. The grey-empty dots represent the first fit with all parameters kept free. The red dots or lines represent the second fit of $\vrot$ and $\vdisp$ after the regularization of the other parameters (Bezier interpolation for $i$ and $\phi$, constant value for $\vsys$, $x_0$ and $y_0$). The cyan-shadowed region and the cyan-dashed lines are the rotation curve (with errors) and the geometrical angles derived by  \citet{deBlok+08} with a 2D fit. For their final rotation curve, de Blok et al. used the coordinates of the centre $(x_0,y_0)=(517,512)$ from \citet{Trachternach+08} and a $\vsys=496.8 \, \kms$, in perfect agreement with our values within the errors.}
\label{fig:n5055}
\end{figure*}

Here we focus on the automatic modeling of the well-known spiral galaxy NGC 5055. This galaxy is a difficult case because it has a prominent warp in the outer disc \citep{Bosma78}. We used THINGS natural-weighted data-cube at a spatial resolution of 10" (about 0.5 $\kpc$, assuming a distance of 10 Mpc) and we run \bba\ by supplying only initial guesses for the inclination and the position angle. \bba\ automatically identifies the galaxy emission in the data-cube, builds the mask, estimates the initial guesses for all the other parameters, performs the first fitting step for each radius with all parameters free and the second fitting step with only $\vrot$ and $\vdisp$ free, after the regularization of the other parameters by Bezier functions (inclination and position angle) or constants. 
The comparison between the observations and the final best-fit model is shown in \reffig{fig:chanmap} through seven representative channel maps (a standard output of \bba).
In \reffig{fig:n5055} we show the resulting values of the fitted parameters in both steps and we compare them with those of \cite{deBlok+08} derived from a tilted-ring fit of the velocity field. Note how the regularization of the geometrical parameters improves the rotation curve, removing for instance the unphysical discontinuities around 28 and 47 kpc. Our final rotation curve in \reffig{fig:n5055} is in very good agreement with that derived by \cite{deBlok+08} from the same data-cube. The main differences arise from kinematic asymmetries between the approaching and the receding halves of the galaxy \citep[see][]{Battaglia+06}. 
While the 2D tilted-ring model applied to the entire disc in an asymmetric galaxy mainly results in a rotation curve usually averaged between the approaching and the receding side, \bba\ always finds the model that has the lowest residuals with respect to the data. This may result in a rotation velocity that is the average between the two halves, or that follows more closely one side rather than the other. This is what happens to our rotation curve in \reffig{fig:n5055}.

The computational time with a regular dual-core laptop for running \bba\ on this THINGS data-cube, sized 1024x1024x87 pixels, is about one day.  High-resolution observations are also suitable to study asymmetries and peculiarities, such as streaming motions, lopsidedness and extra-planar gas. \bba\ always models galaxies with a single rotating disc and it is not designed to handle these peculiarities. In this context, the newest 2D fitting codes might be a more desirable choice, since they guarantee larger possibilities for modeling kinematical and/or geometrical anomalies. The TiRiFiC code also provides for large possibilities of complex modeling, even for those systems where the 2D approach can not be used, like galaxies close to edge-on or having strong spiral arms and thick discs \citep[e.g.][]{Kamphuis+13, Schmidt+14}.

\begin{figure*}
\centering
\includegraphics[width=\textwidth]{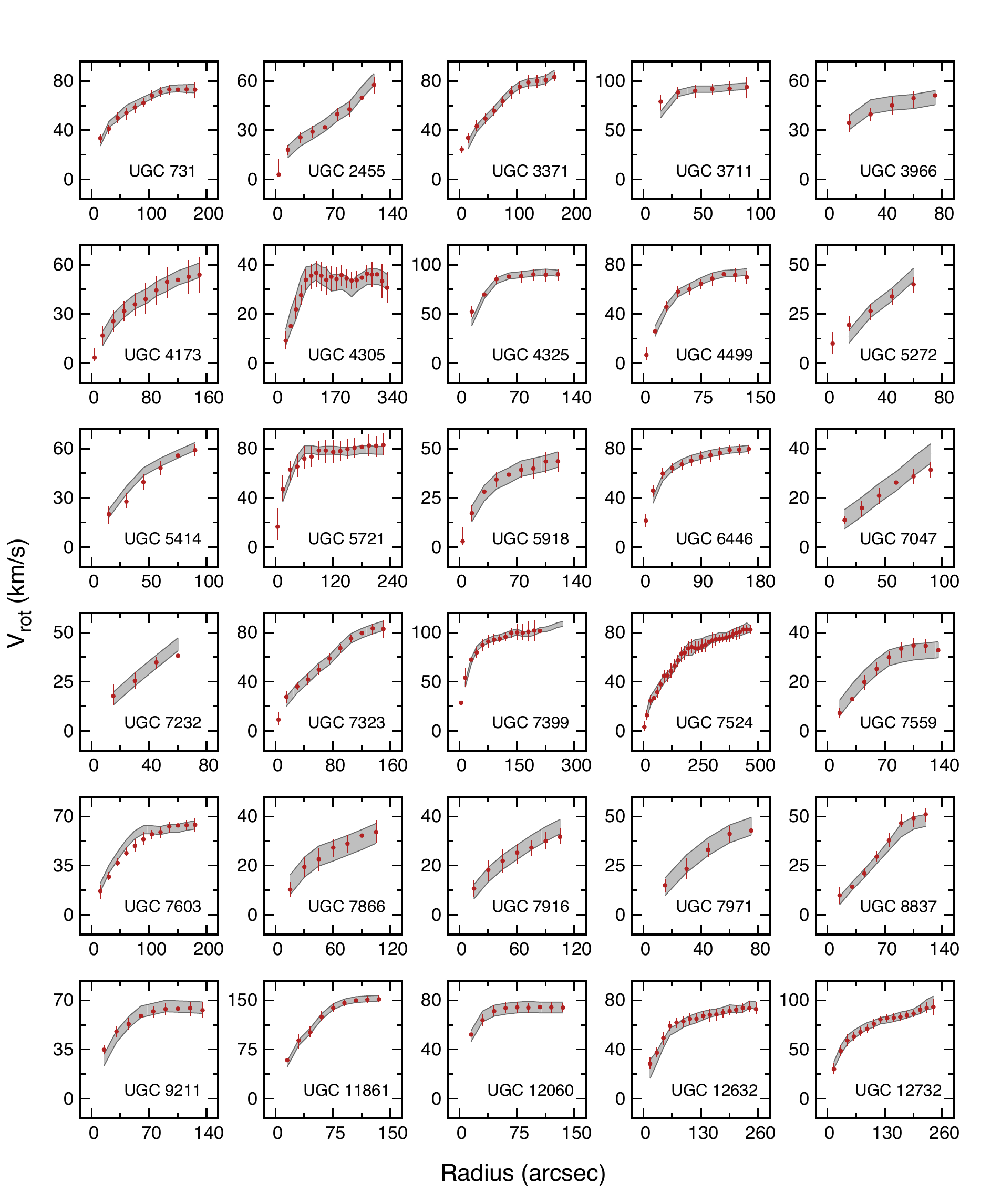}
\caption{Rotation curves of the 30 dwarf late-type galaxies selected from the WHISP sample. The gray-shadowed regions represent the rotation curves from \citet{Swaters99} within the errors, the red dots are the rotation curves derived with \bba . Since Swaters'\ errors are symmetric, his points would lie in the centre of the grey band at the same radii that our points. In UGC 7399, our rotation curve stops earlier because in our data-cube there was no significant emission beyond 175 arcsec.}
\label{fig:whisp}
\end{figure*}

\subsection{Mid-low resolution data and robustness}
\label{sec:whisp}

Only a fraction of local galaxies can be observed with a spatial resolution comparable to that of the THINGS survey.  Most emission-line observations of galaxies, both from radio-interferometers and IFU instruments, currently have less than a dozen resolution elements throughout the entire disc. In these conditions, the beam smearing could heavily affect the derivation of rotation curves with 2D techniques. \bba\ is instead conceived to work with these low-resolution data.

We made a robustness test by using a sample of galaxies from the Westerbork \hi\ survey of Irregular and Spiral galaxies Project \citep[WHISP,][]{vanderHulst+01}. WHISP comprises approximately 350 galaxies in the Local Universe observed in \hi\ with the Westerbork Synthesis Radio Telescope (WSRT). We selected 32 galaxies with published reliable rotation curves \citep{Swaters99}. The sample only contains dwarf late-type galaxies, with $\vrot<100 \kms$,  that usually have a poor spatial resolution and a relatively low signal-to-noise ratio (S/N). For these galaxies, \cite{Swaters99} derived rotation curves using the following procedure that includes a correction for the beam smearing. For each galaxy, a first estimate of the rotation curve was determined by interactively fitting the rotation velocity, position angle and inclination as a function of radius simultaneously to a set of six position-velocity diagrams taken at different angles. This was done by using the \textsc{Inspector} routine in the GIPSY package, which allows to visually inspect different slices and manually tune up the parameters. Centres were fixed to the optical values, the systemic velocities were determined from a tilted-ring fit to the velocity field with the centre fixed. Then, Swaters constructed a 3D model (\textsc{Galmod} in GIPSY) from the estimated parameters and he visually compared it with the observations, iteratively adjusting the input rotation curve and building a new 3D model until the match between model and observation was satisfactory. Such a procedure takes a long time for each galaxy and it may be subjective, but it was the only way to take into account the beam smearing effect and derive reliable rotation curves in these low-resolution data. The approach of \bba\ is analogous, but every step is automatically performed and the best model is quantitatively determined.

We run \bba\ in a semi-blind fashion on the sample of 32 dwarf galaxies. We used 30" smoothed data-cubes and we set a ring width of 15", following \cite{Swaters99}. The systemic velocities and the coordinates of the centre of the galaxies were automatically estimated through the source finding algorithm and fixed to those values. In each case, values perfectly compatible with those of \cite{Swaters99} were obtained: the maximum deviations from Swaters values are $\Delta\vsys<2\,\kms$ and $\Delta x_0 \simeq \Delta y_0 < 1 \, \mathrm{pixel} = 10$". Global initial estimates for rotation velocities, velocity dispersions and thicknesses of the discs were set to $\vrot=50 \kms$, $\vdisp = 8 \kms$ and $z_0 = 200 \pc$ for all galaxies, while the initial inclinations and the position angles were taken from Tab. A1, Chap. 4 of \cite{Swaters99}. We used the $\abs{\,M-D\,}$ residuals with a $\cos(\theta)$ weighting function. Masks were built by smoothing the data by a factor 2 and considering only those regions with flux $>3$rms, being rms the root mean square (noise) of the smoothed data-cube. \bba\ performed a first step by fitting $\vrot $ , $\vdisp$, $i$ and $\phi$ and a second step by fitting only $\vrot $ and $\vdisp$ and fixing $i$ and $\phi$ to a 2$^\mathrm{nd}$ degree polynomial function. The execution time is less than a minute per galaxy on a regular laptop.

A comparison between Swaters' rotation curves and ours is shown in \reffig{fig:whisp}. Position-velocity diagrams (output of \bba) along major and minor axes are shown in Appendix A (available as online material). In general, the agreement is very good. Most differences in the rotation velocities can be attributed to asymmetries in the kinematics between the receding and approaching halves of the galaxies. It is interesting to notice that in some galaxies our rotation curves rise more steeply than Swaters' in the inner regions (e.g.\ UGC5272, UGC 6446, UGC 9211). Since the main effect of the beam smearing is to reduce the velocity gradients in the rising part of the rotation curve, it is possible that the correction manually made by  \cite{Swaters99} was not sufficient in those cases.

Out of 32 data-cubes, \bba\ failed in determining acceptable models for 4 galaxies, either not converging or deriving wrong kinematics. Two cases (UGC 3966, UGC 8837) can be attributed to a wrong fit of the inclination in the first step and fixing it to the initial value led to a good model (see \reffig{fig:whisp}). 
In the last two cases, UGC 7690 and UGC 8490, we could not make the code working manually neither. The first galaxy is very faint and the fit is hampered by the noise. The galaxy UGC 8490 has a huge warp in inclination and the algorithm tries to reproduce it by varying the velocity rather than the inclination angle. The final model looks good, but the rotation curve is  probably unphysical. 
In addition, from the inspection of the position-velocity diagrams along the minor axis, it turned out that the automatic estimate of the centres was slightly inaccurate in five galaxies. Putting the optical centres manually led to better models, even though the rotation curves did not change significantly. Overall, 94\% of the galaxies were accurately modeled by \bba.

From this test emerges that \bba\ is able to derive reliable kinematics in low-resolution and noisy data-cubes. We remind that we run \bba\ in a almost blind execution, since the only information we supplied to the code were the initial guesses for the inclination and the position angles. Of these, the inclination is especially critical as it may be often unsuccessfully estimated by the code. 
Improvements in the initial parameters estimate algorithm will be considered in the next releases. 
We stress however that the careful inspection of the outputs (position-velocity diagrams and model cubes) does clearly single out cases where the fit was not successful.
Our test on the WHISP data reveals that the success rate is very high. This, combined with the very low computational time needed to fit these low-resolution galaxies ($\lesssim$ 1 minute) are key features for the application of \bba\ to the upcoming large \hi\ surveys. Indeed, already planned \hi\ survey, such as WALLABY and DINGO with ASKAP \citep{Johnston+08}, LADUMA with MeerKAT \citep{Booth+09} and WNSHS with WSRT/APERTIF \citep{Verheijen+08}, are expected to observe thousands of galaxies with spatial resolution comparable to the WHISP galaxies.

\subsection{Very low-resolution data}
\label{sec:verylow}

In this section we show the effect of the beam smearing on the derivation of rotation curves from 2D and 3D analysis, going down to extremely low spatial resolution. 

\begin{figure}
	\begin{subfigure}
	\centering
	\includegraphics[width=0.475\textwidth]{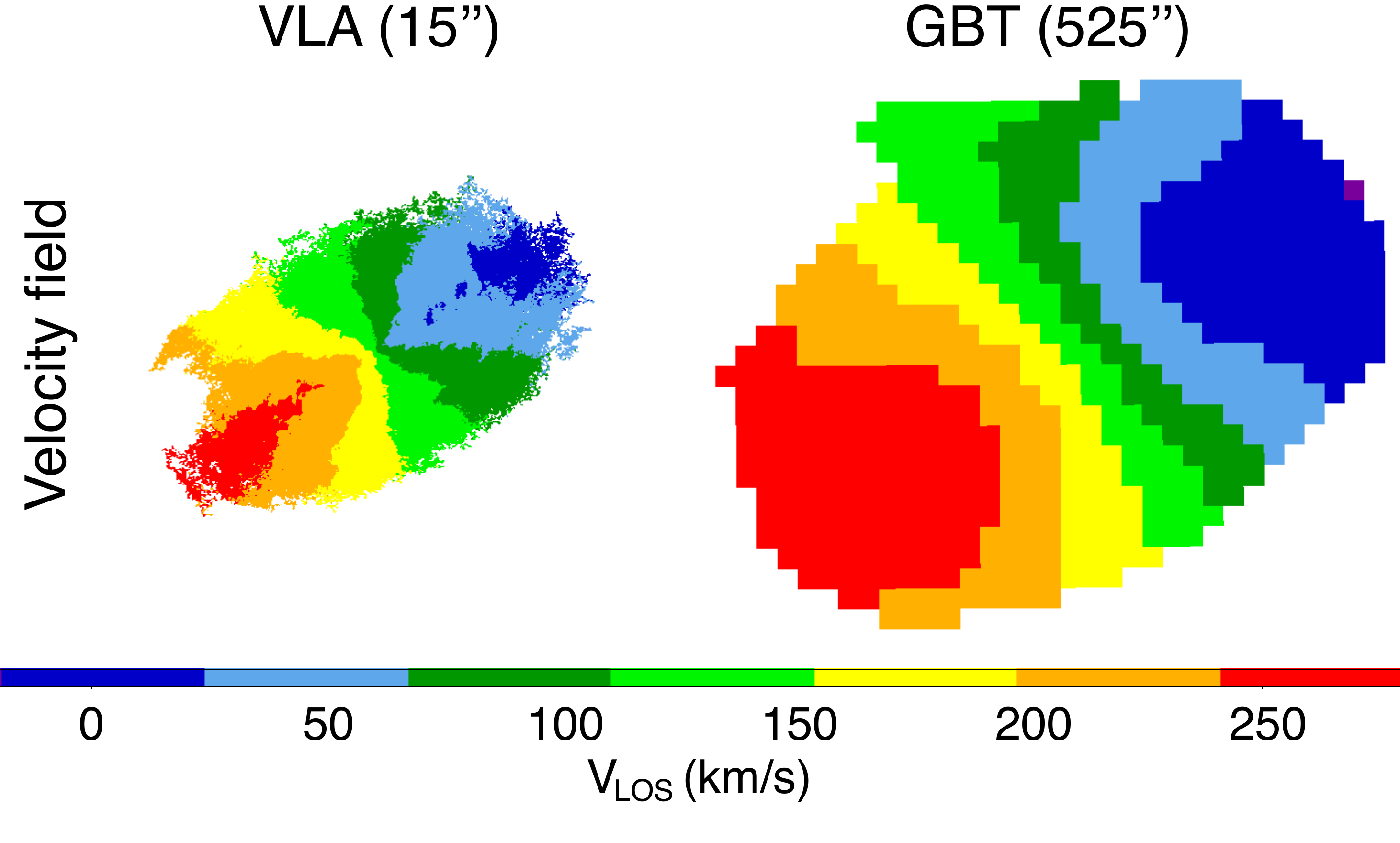}
	\vspace{0.1cm}
	\end{subfigure}
	\begin{subfigure}
	\centering
	\includegraphics[width=0.475\textwidth]{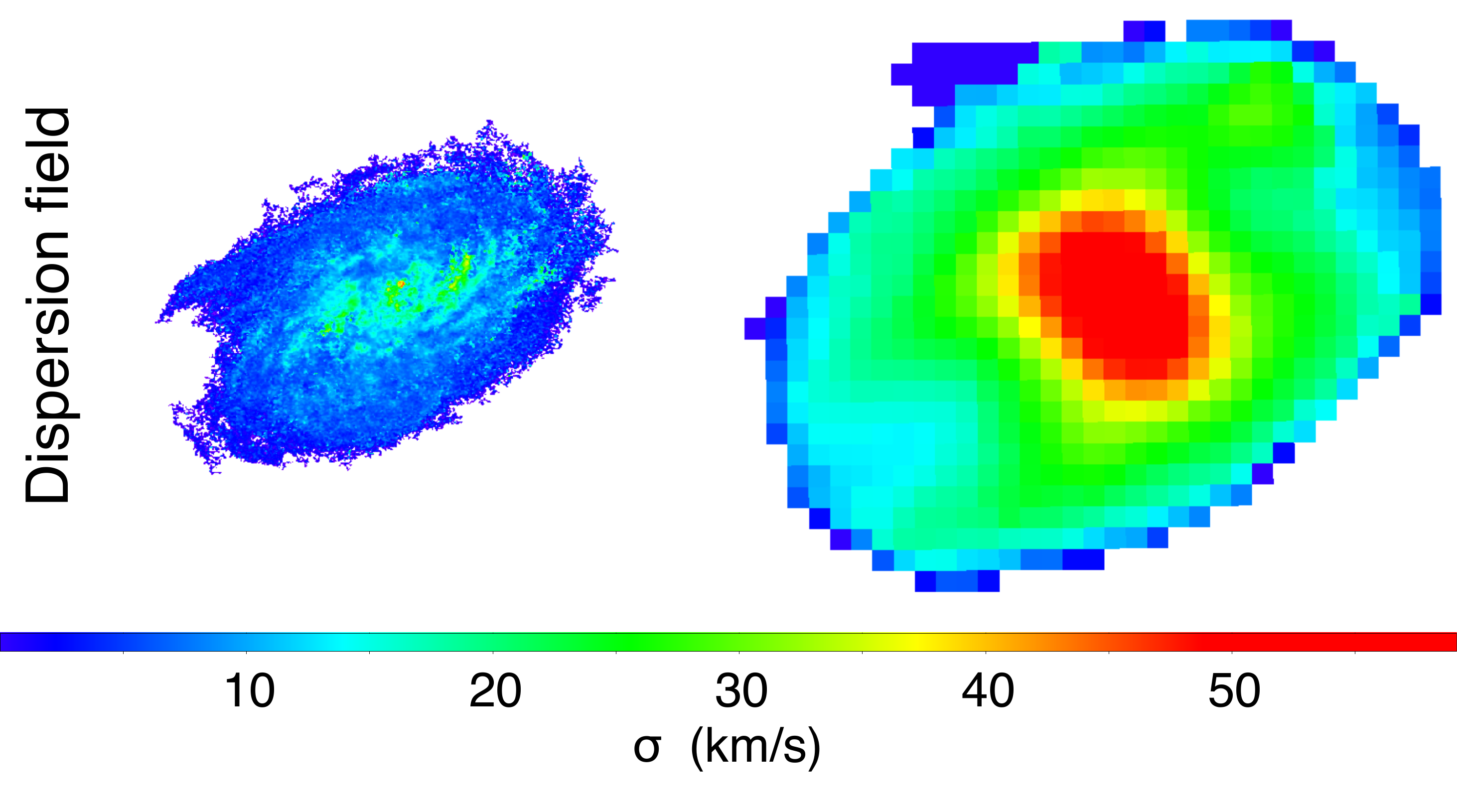}
	\end{subfigure}
\caption{Velocity fields ($1^\mathrm{st}$ moment) and velocity dispersion fields ($2^\mathrm{nd}$ moment) for NGC 2403 derived from high \citep[\emph{left},][]{Fraternali+02} and low resolution  \citep[\emph{right},][]{deBlok2+14} data-cubes. Images are on the same spatial and velocity scale. Note the dramatic effect of the beam smearing: the velocity field at high resolution, showing the typical traits of a flat rotation curve, turns into a nearly solid-body pattern at low resolution, especially in the inner parts. Velocity dispersions increase by a factor 3-4 throughout the whole disc.}
\label{fig:n2403maps}
\end{figure}

\begin{figure*}
\centering
\includegraphics[width=\textwidth]{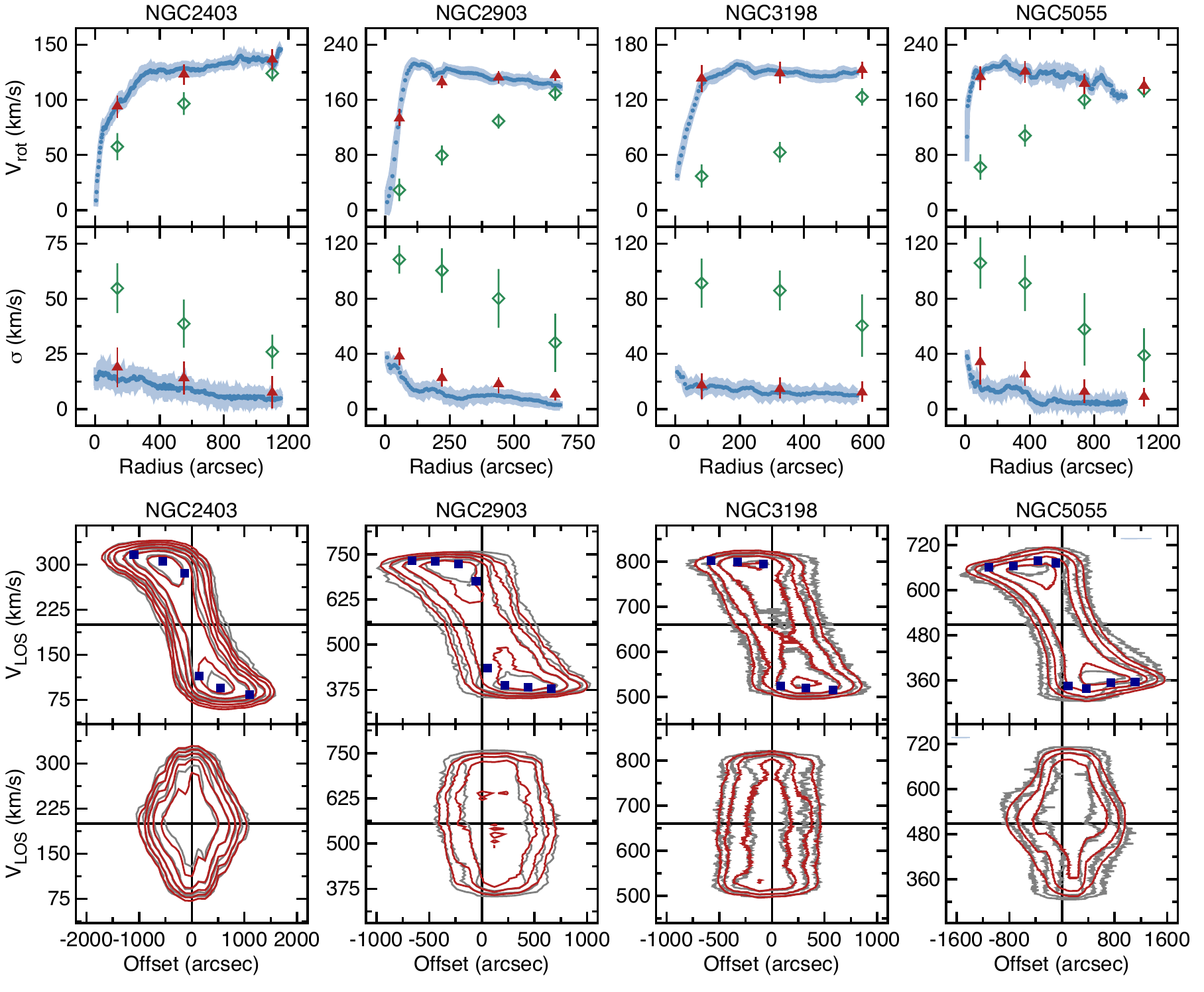}
\caption{\emph{Upper panels}: velocities (\emph{top}) and velocity dispersions (\emph{bottom}) for NGC 2403, NGC 2903, NGC 3198 and NGC 5055. Blue dots were derived from THINGS high-resolution (6") data-cubes with \bba . Cyan regions represent the errors. Red triangles are the fit with \bba\ in single-dish data, the green open diamonds are the results obtained from the 2D maps (\rotcur\ on velocity fields and \textsc{Ellint} on dispersion fields). \emph{Lower panels}: the correspondent position-velocity diagrams along major (\emph{top}) and minor (\emph{bottom}) axes. Data are represented in grey, models in red, rotation curves as blue square dot.}
\label{fig:lowres}
\end{figure*}

We run both \bba\ and \rotcur\ on a small sample of nearby galaxies observed in \hi\ with single-dish telescopes. We selected 4 galaxies (NGC 2403, NGC 2903, NGC 3198 and NGC 5055) for which we have both very high resolution \hi\ data \citep[][and THINGS]{Fraternali+02} and single-dish observations. Effelsberg data for NGC 3198 and NGC 5055 were kindly provided by B. Winkel \citep*{Winkel+12,Winkel+12_2}. NGC 2403 \citep{deBlok2+14} and NGC 2903 (Pisano et al., in prep.) where observed with the Green Bank Telescope (GBT) and kindly supplied by D.\ J.\ Pisano. We compared high-resolution rotation velocities and velocity dispersions obtained with \bba\ with the low-resolution ones derived both with 2D and 3D approaches. Typical spatial resolution is 6" for the high-resolution data and 650" (Effelsberg) or 525" (GBT) for the low-resolution data, which means that these galaxies are barely resolved. High-resolution rotation curves and dispersions were obtained with \bba\ from the natural-weighted THINGS data-cubes as described in \refsec{sec:hires}. Low-resolution velocity and dispersion fields were derived as $1^\mathrm{st}$ and $2^\mathrm{nd}$ moment maps, respectively (see e.g.\ NGC 2403, \reffig{fig:n2403maps}). In the 2D approach, rotation curves were derived with \rotcur\ fitting only $\vrot$ and fixing the other parameters to the high-resolution values, 2D velocity dispersion profiles were obtained by taking the average value along the rings on the dispersion fields. No correction for the beam smearing was performed. \bba\ was run by fitting only $\vrot$ and $\vdisp$, except for NGC 5055, where we kept free also the position angle to trace the outer warp. For NGC 2403, we used only the receding half of the disc, since the approaching half is contaminated by \hi\ emission from the Milky Way. In \reffig{fig:lowres} we show the resulting rotational velocities, velocity dispersions and the comparison between the models and the data through the position-velocity diagrams along the major and the minor axes. As expected, a 2D approach is not suitable for data at these resolutions, since the beam smearing significantly affects the derivation of the 2D maps from the data-cubes.
Beam smearing flattens the gradients in the velocity profiles and turns rotation velocity into apparent high velocity dispersion, as it clearly appears from the maps in \reffig{fig:n2403maps}. Such a degeneracy between $\vrot$ and $\vdisp$ is broken in \bba , because the beam smearing effect is taken into account in the convolution step (\refsec{sec:conv}). Unlike the 2D approach, \bba\ recovers correct rotation velocities for all these galaxies in every location of the disc. The differences between the low-resolution and the high-resolution velocities are within the errors in all cases. Even more remarkably, \bba\ returns low values for the intrinsic velocity dispersions of the gas that are fully comparable with the correct values (bottom panels of the upper plots). An inspection of the position-velocity diagrams along the minor axis (bottom panels of the lower plots) should give an idea of how the line broadening in these data is fully dominated by instrumental effects.

These results show that \bba\ is a powerful tool to study the kinematics of galaxies even in very low-resolution data (2-3 resolution elements across the whole disc), where the standard 2D approach fatally fails. Above all, \bba\ is almost always able to describe the correct shape of the rotation curves and it becomes extremely robust when its outputs are visually checked by the user.
A particularly interesting application concerns the Integral Field Spectrographs (IFS) data. Instruments like SINFONI \citep{Eisenhauer+03}, KMOS \citep{Sharples+08} and MUSE \citep{Bacon+10} on the Very Large Telescope (VLT) can observe line emission, such as $\ha$, N and O forbidden lines, in galaxies up to redshift about 2.5, with a resolution similar to the observations showed in this section. Running \bba\ on those data-cubes is a very challenging task because, in addition to the poor spatial resolution, these observations have a low spectral resolution (channel widths of 30-40 $\kms$ compared to the few $\kms$ of \hi\ data) and 
very low S/N. Nevertheless, a 3D approach should be heartily recommended in order to take advantage of the full information available in these data-cubes.

\section{Testing $^{\scriptstyle \mathbf{3D}}$Barolo's limits}
\label{sec:mock}

\begin{figure}
\centering
\includegraphics[width=0.48\textwidth]{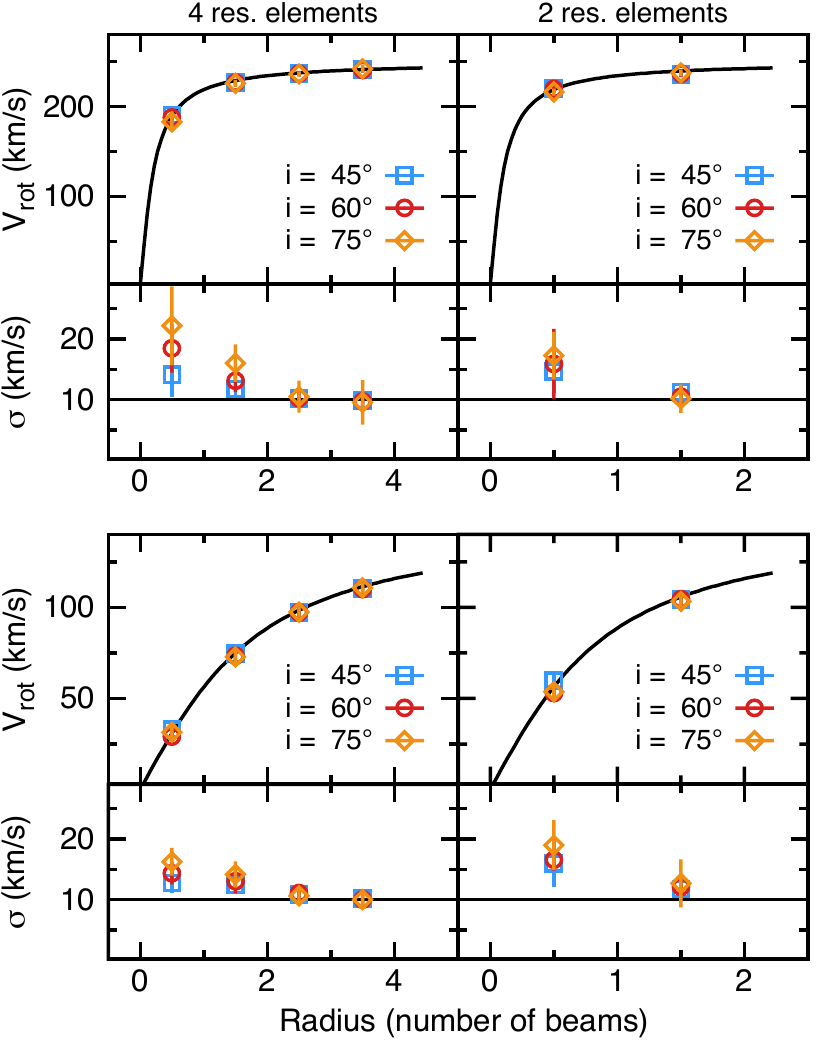}
\caption{Rotation velocity and velocity dispersions derived with \bba\ from models with flat (\emph{top}) and slowly-rising (\emph{bottom}) rotation curves. Here we show the lowest resolution models, i.e. 100" (\emph{left}) and 200" (\emph{right}), and three inclinations (45$\de$, 60$\de$ and 75$\de$). Models are the black thick lines. At these resolutions, we run \bba\ with just $\vrot$ and $\vdisp$ free. }
\label{fig:inc_var}
\end{figure}

The goodness of a fit is mainly determined by the combination of four factors: the inclination of the galaxy disc, the spatial resolution, the spectral resolution and the S/N. We used mock galaxies to find some significant thresholds to these quantities and test under what conditions \bba\ may have problems in deriving a reliable kinematic model.

We built initial artificial data-cubes with a resolution of 12" (FWHM), a channel width of 7.5 $\kms$ and no noise. The pixel size is 3", thus the beam covers an area of about 18 pixels. 
The galaxy models have a maximum radius of 400", i.e.\ about 33 spatial resolution elements per side.
This configuration could be a typical observation of a nearby galaxy with a modern interferometer, like the JVLA. We set global parameters for $x_0$, $y_0$, $\phi$, $\vsys$ and $z_0$, an exponential gas density profile and a constant dispersion field of 10 $\kms$. The initial inclination is $i=60^\circ$. We assumed rotation curves shaped as $V_\mathrm{rot}\,(R)= 2/\pi\, V_0 \arctan \,(R/R_0)$, being $V_0$ the asymptotic circular velocity and $R_0$ the turnover radius, i.e.\ the transitional point between the rising and flat part of the rotation curve. We made a model with a steeply-rising plus flat rotation curve ($V_0=250 \,\kms$, $R_0=20$") and a model with slowly-rising solid body-like rotation curve ($V_0=150 \,\kms$, $R_0=150$"). We progressively degraded the artificial data-cubes and we try to get back the input rotation curves using \bba.

We first reduced the spatial resolution by smoothing the initial data-cubes down to 200" (preserving the number of pixels per beam), that is 2 resolution elements per side for the model galaxies. With a large number of velocity channels and no noise, \bba\ is able to recover the correct rotation curve even at the lowest resolution. 
This could be the case of a single-dish observation as those in \refsec{sec:verylow}. For comparison, we also derived the rotation curves running \rotcur\ on the Gaussian velocity fields, fitting only the rotation velocity. Both approaches recover an almost perfect rotation curve when the galaxy has more than 10 resolution elements, but, below this limit, \rotcur\ increasingly underestimates the rotation velocity, especially in the inner regions of the model with flat rotation curve, whereas \bba\ can successfully determine it. With the 2D approach the relative errors with respect to the actual rotation curve are up to 60\% in the lowest resolution and in the inner parts, whereas with the 3D approach the errors are confined to a few percent at all resolutions and over the entire disc. The derived rotation curves obtained at 100" and 200"  (4 and 2 resolution elements respectively) are shown in \reffig{fig:inc_var} (red circles).

Next, we studied the effect of the inclination angle on \bba's accuracy, in particular on the fit of $\vrot$ and $\vdisp$. We let the inclination of the model galaxies varying from nearly face-on to nearly edge-on and fitted these models by fixing the inclination. In \reffig{fig:inc_var} we show the recovered rotation curves and velocity dispersions for the lowest resolution models, namely two and four resolution elements per side, and for three significant inclination angles, i.e.\ 45$\de$, 60$\de$ and 75$\de$. 
The rotation velocity is well recovered at any inclination both in models with flat and solid-body rotation curve, although in the flat model for $i>75\de$ the inner points of the rotation curves start to be underestimated. 
The velocity dispersion of the inner point can be overestimated in some cases by a factor up to about 2 but with large error bars, the other points are recovered within a few $\kms$.

Overall, if the inclination is known, \bba\ can derive the correct rotation curve and can disentangle between rotation and velocity dispersion at almost every inclination and even for data at very low spatial resolution where the 2D approach can not be used. 
When the inclination is not known, the code can fit it together with the kinematical parameters, but  this requires some care. 
The inclination is the thorniest parameter to deal with, and running \bba\ with a completely unknown inclination is not advisable since the code can easily converge to a local minimum close to the initial inclination. 
The initial guess for the inclination is therefore essential for the goodness of the fit and it can be either supplied by the user (e.g. optical values) or estimated by \bba. 
Our tests show that the algorithm for the initial guesses (see \refsec{sec:addfeat}) returns good estimates (within a few degrees) of the global inclination in most cases regardless of the spatial resolution when $45\de \lesssim i \lesssim 75\de$. 
Thus, in general for $i\gtrsim45\de$ one should always be able to obtain acceptable kinematical fits (note that above $75\de$, errors in the inclination have little impact in the rotation velocity).
However, for $i\lesssim45\de$ rotation curves may become progressively more uncertain due to the smaller rotational component along the line of sight and the larger impact of inclination errors. 
This is a problem for any fitting algorithm.
Finally, we note that the inclination is degenerate with the disc thickness and we expect this effect to be important especially for dwarf galaxies.
In the future, we will consider to include a self-consistent treatment (assuming the hydrostatic equilibrium of the gas) for the disc thickness in \bba.

\begin{figure}
\centering
\includegraphics[width=0.48\textwidth]{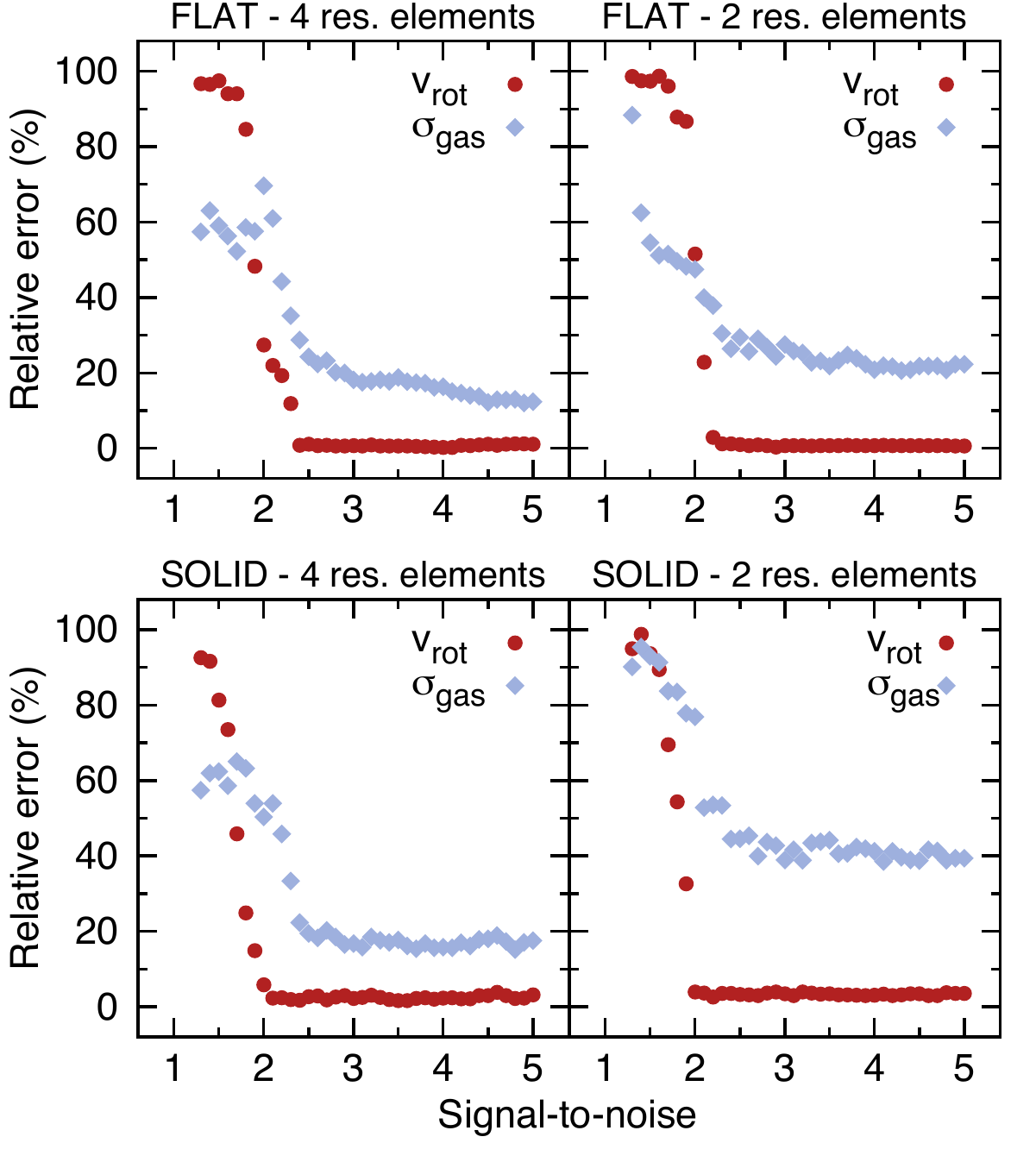}
\caption{Relative percentage errors for the fitted $\vrot$ (red circles) and $\vdisp$ (cyan diamonds) as a function of the average S/N at fixed $i=60\de$. We show the results for the flat (\emph{top}) and solid-body model (\emph{bottom}) at the lowest resolutions, i.e.\ 100" (4 resolution elements, \emph{left}) and 200" (2 resolution elements, \emph{right}). The code becomes unreliable below a S/N $\lesssim2-2.5$.}
\label{fig:sn_var}
\end{figure}

We also tested the effect of the noise on the derivation of rotation velocities and velocity dispersions. We focused on $i=60^\circ$, used a constant surface density profile for the galaxy models and added a progressively higher Gaussian noise to the artificial data-cubes with two and four resolution elements. Relatively low S/N and less than 10 resolution elements characterize ALMA data of high-z galaxies and are expected for most data of the future \hi\ surveys. 
We calculated the relatives percentage errors of the fitted $\vrot$ and $\vdisp$ with respect to the true values. In \reffig{fig:sn_var} we show the average errors over the entire disc as a function of the average S/N. 
The performance of \bba\ remains consistently high whenever the signal to noise of the average emission in the ring is S/N $>2$.
Below this limit, \bba\ may consider portions of the background as galaxy emission, leading to a wrong fit of the circular velocity and velocity dispersion, which can be either overestimated or underestimated without systematic changes in slope or shape. Further tests in the inclination range $45\de<i<75\de$ did not significantly change such a threshold and the trends showed in \reffig{fig:sn_var}. 
We stress that these tests were run with the default masking options and that a manual fine-tuning of the mask might lead to good results even at slightly lower signal-to-noise ratios.

Finally, we reduced the spectral resolution in the noisy artificial data-cubes. Low spectral resolutions characterize, for instance, IFUs data (like KMOS or MUSE), which typically have channel widths of 30-40 $\kms$ for each emission line, i.e.\ observed galaxies can span less than ten channels. Increasing the channel width dramatically lowers the number of data points that \bba\ can use to constrain the best model. From our test emerges that the number of channels that guarantees a good fit at low S/N varies between 8 and 12, depending on the spatial resolution.

In conclusion, \bba\ can work even with observations at very low spatial/spectral resolution and low S/N and in a wide range of galaxy inclinations. These factors influence together the goodness of the model. In an extreme case of a galaxy with just a couple of resolution elements per side, it would be advisable that the source is detected at a S/N $\gtrsim$ 3 and over a dozen channels or more.

\section{Conclusions and future prospects}
\label{sec:conclusions}
In this paper, we presented \bba, a new software for the fitting of 3D tilted-ring models to emission-line observations, and we showed several examples of applications, from high-resolution to very-low resolution data-cubes. The main purpose which led us to develop \bba\ is to have a tool to derive the kinematics of galaxies using a simple rotating disc model, feasible to be used at a wide range of resolutions without being affected by instrumental effects. At high resolution, \bba\ works as well as the 2D traditional approach, but it is more computational expensive and it does not yet allow to study the peculiarities of galaxies, such as non-circular motions. \bba\ reaches its best performance in deriving reliable rotation curves from low-resolution data, down to barely resolved galaxies, where the 2D approach totally fails because of the beam smearing. Moreover, the robustness and the possibility of identifying sources and estimating the initial conditions for the fit make \bba\ a tool that can be automatically run on large amounts of data. 

\bba\ can operate on all emission-line data. In particular, suitable observations are the \hi-line, CO and C$^+$ lines from sub-millimeter interferometers, such as the Plateau de Bure Interferometer (PdBI) or ALMA, and  optical/IR recombination lines from the last generation of IFUs. These instruments are providing the extraordinary opportunity to study the evolution of the kinematics and the dynamics of galaxies throughout the Hubble time.
Several recent studies have used these data to investigate how the earliest disc galaxies differ from local galaxies through different analysis, like the V/$\sigma$ ratio, which measures the dynamical support given by rotation \citep[e.g.][]{ForsterSchreiber+06,Mancini+11}, the Toomre Q parameter \citep[e.g.][]{Genzel+11} or the turbulence in the disc \cite[e.g.][]{Green+10}.
Running \bba\ on these observations will give the opportunity to derive reliable rotation curves at high-z and break the degeneracy between rotation velocity and velocity dispersion down to very low spatial resolutions.

Overall, the kinematics is the starting point to investigate several properties of disc galaxies, including but not limited to the dynamics, the matter distribution and the evolution of scaling relations  \citep[e.g.][]{Epinat+09,Miller+12}. 
The ability to quickly derive rotation curves will be more and more important in the future given the increasing number of emission-line observations available from various astronomical facilities.

\section*{Acknowledgements}
We thank the anonymous referee for helpful comments and suggestions. We thank Benjamin Winkel, George Heald and D.\ J.\ Pisano for providing their proprietary \hi-data. This research made use of the WHISP and THINGS data sample and of some parts of the Duchamp code, produced at the Australia Telescope National Facility, CSIRO, by Matthew Whiting. We acknowledge financial support from PRIN MIUR 2010-2011, project \vir{The Chemical and Dynamical Evolution of the Milky Way and Local Group Galaxies}, prot. 2010LY5N2T.

\bibliography{biblio}

\appendix
\section{P-V diagrams of WHISP galaxies}
In this appendix we show the comparison between our best models and the observations for 30 dwarf galaxies selected from the WHISP sample (see Section 3.2). For each galaxy we show  slices along  major and minor axes (outputs of \bba). Data are in grey and cyan (negative contours) and the model in red. Dark-blue square are the derived rotation curves.

\begin{figure*}
\includegraphics[width=\textwidth]{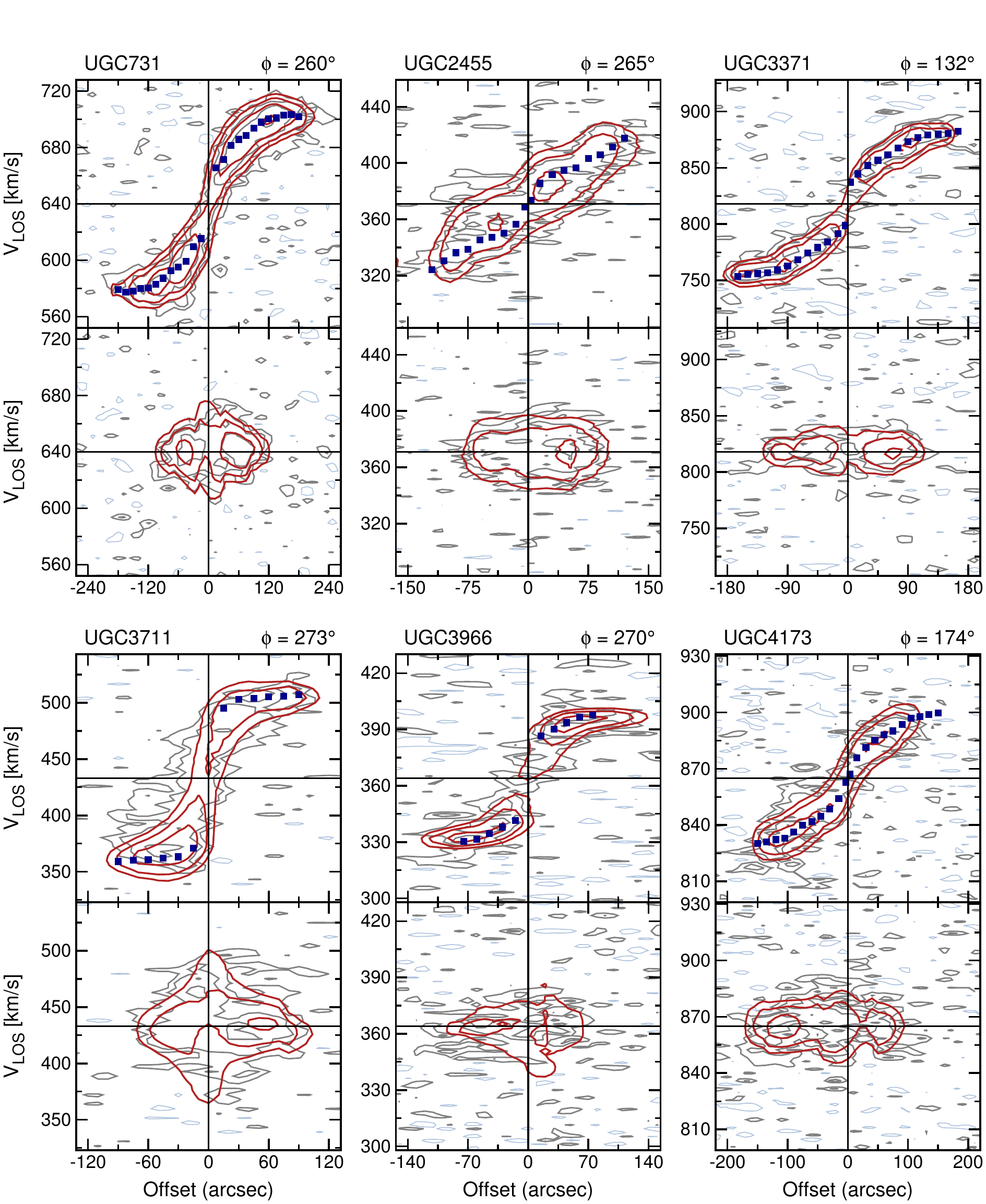}
\end{figure*}

\begin{figure*}
\includegraphics[width=\textwidth]{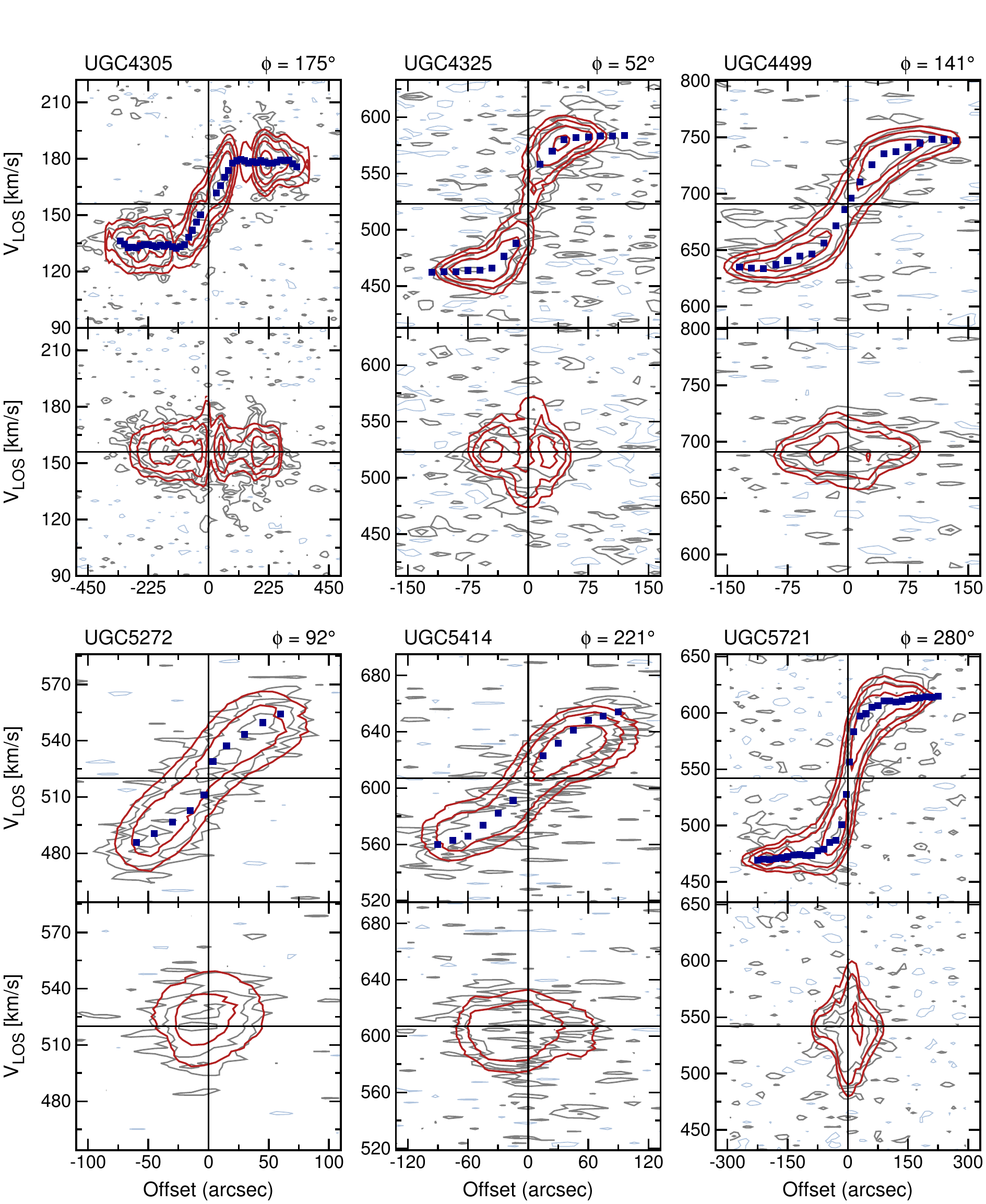}
\end{figure*}

\begin{figure*}
\includegraphics[width=\textwidth]{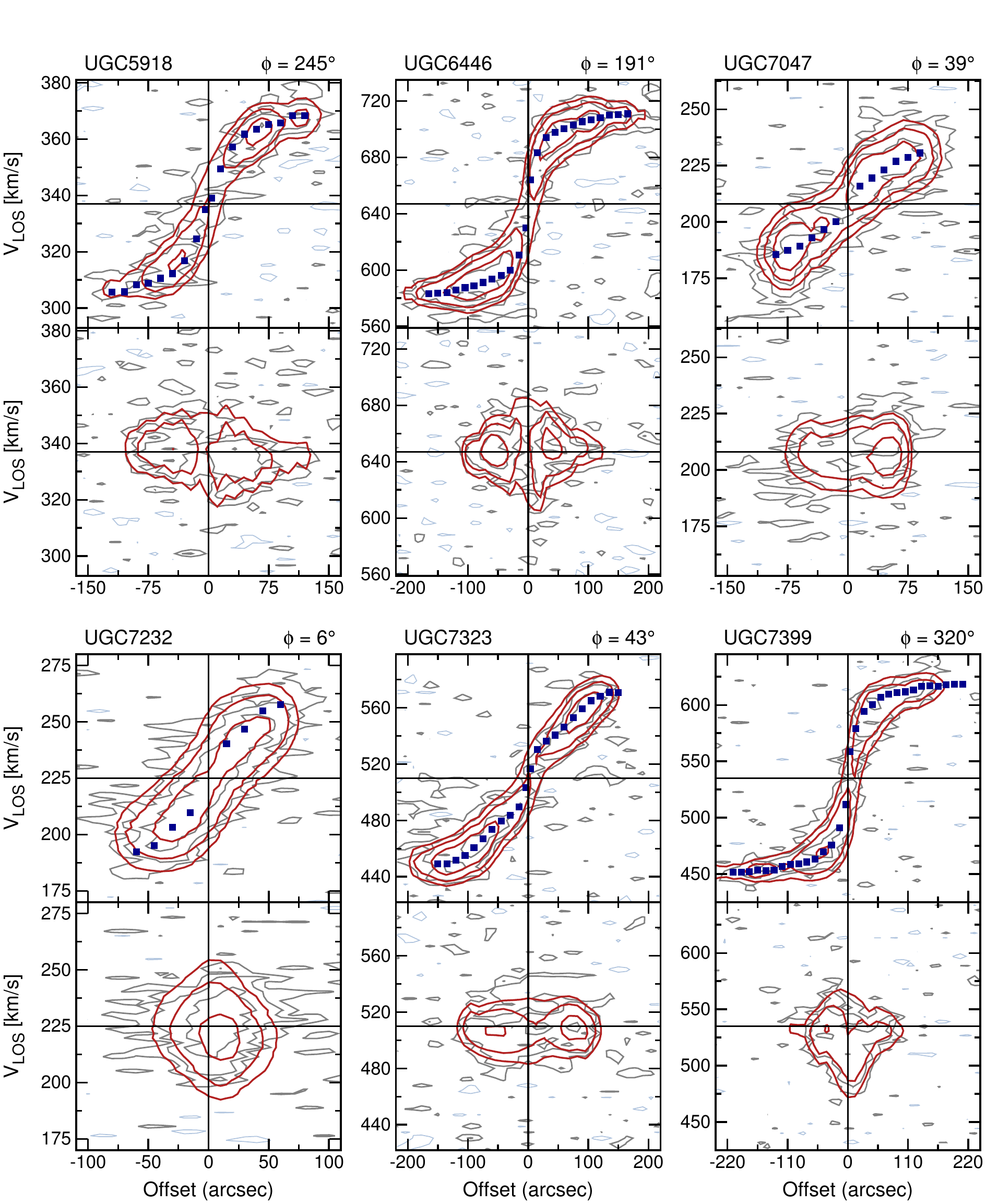}
\end{figure*}

\begin{figure*}
\includegraphics[width=\textwidth]{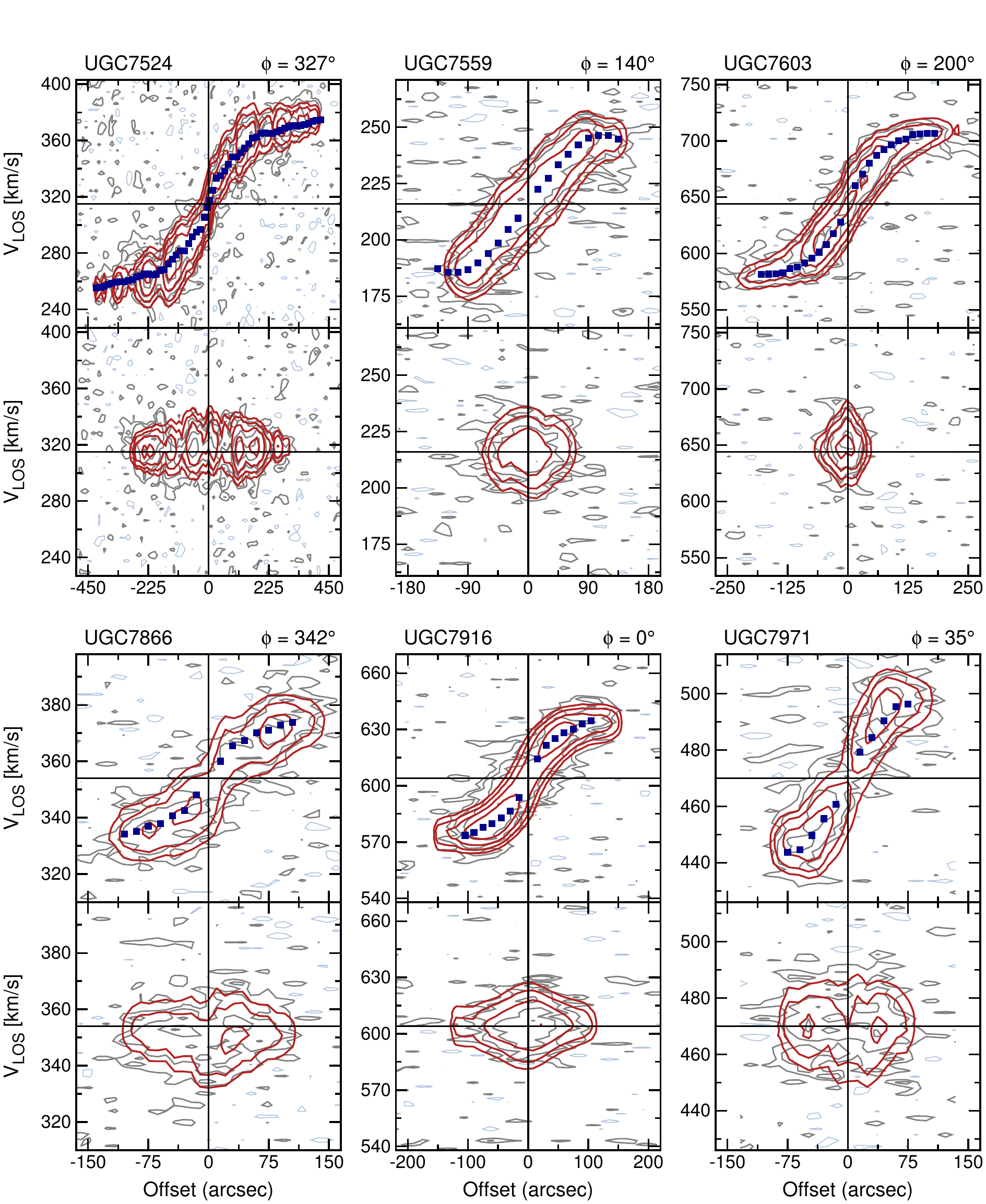}
\end{figure*}

\begin{figure*}
\includegraphics[width=\textwidth]{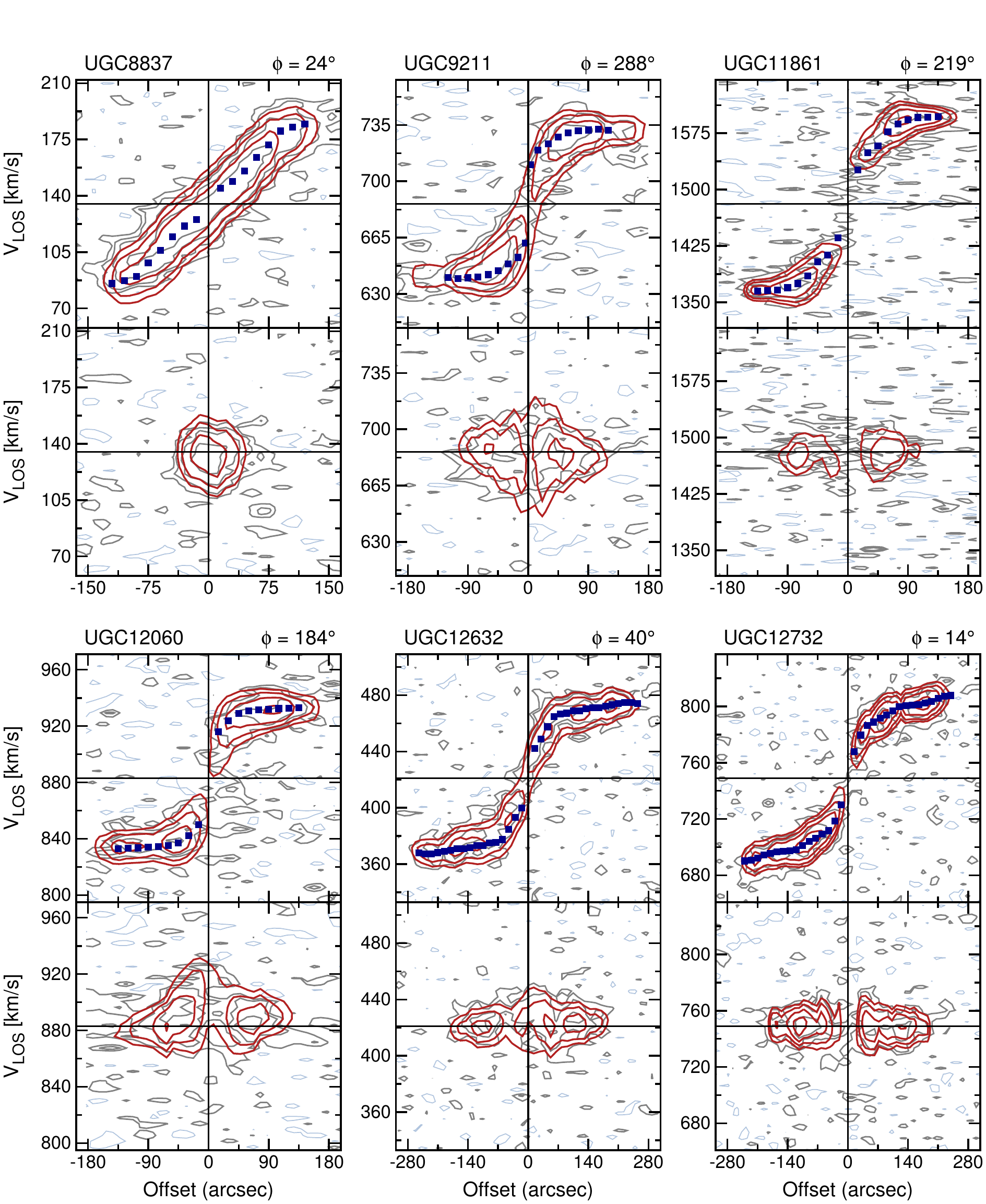}
\end{figure*}

\end{document}